\documentclass{ar-1col-S2O}

\usepackage[comma]{natbib}
\usepackage{url}
\usepackage[hidelinks=true]{hyperref}

\jname{Xxxx. Xxx. Xxx. Xxx.}
\jvol{AA}
\jyear{YYYY}
\doi{10.1146/((please add article doi))}

\usepackage[utf8]{inputenc}
\usepackage[T1]{fontenc}
\usepackage{amsmath, amssymb}
\usepackage{xspace}
\graphicspath{{./}, {figures/}}
\newcommand{\reffig}[1]{\textbf{Figure \ref{#1}}}
\newcommand{\refeq}[1]{Equation \ref{#1}}

\newcommand{\refsec}[1]{Section \ref{#1}}
\newcommand{\msun}{{\rm M}_{\odot}}
\newcommand{\lsun}{{\rm L}_{\odot}}
\newcommand{\rsun}{{\rm R}_{\odot}}

\newcommand{\kms}{{\rm km\,s^{-1}}}
\newcommand{\myr}{{\rm Myr}}

\newcommand{\arepo}{\mbox{\textsc{Arepo}}\xspace}

\newcommand{\eg}{e.g.\@\xspace}
\newcommand{\cf}{c.f.\@\xspace}
\newcommand{\ie}{i.e.\@\xspace}
\begin{document}

% Page header
\markboth{Schneider}{Stellar Mergers}

% Title
\title{Theory, Simulations and Observations of Stellar Mergers}

%Authors, affiliations, address.
\author{Fabian R.~N. Schneider,$^{1,2}$
\affil{$^1$Heidelberger Institut f{\"u}r Theoretische Studien, Heidelberg, Germany; email: fabian.schneider@h-its.org}
\affil{$^2$Zentrum f{\"u}r Astronomie der Universit{\"a}t Heidelberg, Astronomisches Rechen-Institut, Heidelberg, Germany}}

%Abstract (225 words max)
\begin{abstract}
Stellar mergers are responsible for a large variety of astrophysical phenomena. They form blue straggler stars, give rise to spectacular transients, and produce some of the most massive stars in the Universe. Here, we focus on mergers from binary evolution and stellar collisions but do not cover mergers involving compact objects. We review how mergers come about, explain the physics and outcome of the merger process, discuss the evolution and ultimate fates of merged stars, and relate to observations. % 80 words
\begin{itemize}% 3-5 items max with key points
\item Mergers of main-sequence stars often fully rejuvenate and

have interior structures similar to genuine single stars. % 17 words

\item Contrarily, mergers involving post-main-sequence stars can have

interior structures that cannot be achieved by single-star evolution.

Such merged stars may become long-lived blue supergiants that

can explode in SN~1987A-like events, interacting and super-

luminous supernovae, ultra-long gamma-ray bursts or collapse 

into very massive black holes. These black holes may even 

populate the pair-instability-supernova black-hole mass gap. % 59 words

\item Strong magnetic fields are produced in stellar mergers. Merged 

stars may thus be at the origin of some magnetic OBA stars and 

their descendants, highly magnetic white dwarfs and neutron stars. % 29 words

\item Initially, stellar merger products rotate rapidly, but there are 

several mechanisms that can quickly spin them down. Hence, 

merged stars may be rather slow rotators for most of their evolution. % 30 words

\end{itemize}
\end{abstract}

%Keywords, etc.
\maketitle

%Table of Contents
\tableofcontents

%
% Intro
%
\section{INTRODUCTION}\label{sec:intro}

Stellar mergers first came into the spotlight of astrophysical research when they were considered to power the energy release in the dense extragalactic nuclei of quasi-stellar radio sources \citep[quasars; see, \eg][]{woltjer1964a, gold1965a, vandenbergh1965a, spitzer1966a, colgate1967a}. The proposed mechanisms for powering quasars in this way included the release of energy from stellar collisions, nuclear burning induced by high post-collision temperatures in the merged stars, and high supernova (SN) rates maintained by continued collisions. \citet{mathis1967a} and later \citet{deyoung1968a} conducted the first 1.5D `mirror' merger models of twin stars to better understand the transients produced by stellar collisions and the associated mass loss. It quickly became clear that stellar collisions cannot power quasars. These early works on collisions were followed by 2D computations of \citet{seidl1972a}, and finally the first 3D smoothed-particle hydrodynamic (SPH) simulations by \citet{benz1987a}. 

Today, stellar mergers are central to much of astrophysics. The mergers of white dwarfs (WDs) can result in R~Coronae Borealis stars and Type~Ia supernova (SN) explosions \citep[\eg][]{webbink1984a, ruiter2025a} that led to the discovery of the accelerated expansion of our Universe \citep{riess1998a, perlmutter1999a}, and the detection of gravitational waves from merging neutron stars (NSs) and black holes (BHs) has opened the window to gravitational wave astronomy \citep{abbott2016a, abbott2017b}. Mergers of non-compact objects and common envelope (CE) phases give rise to luminous red novae (LRNe), defined by the prototype transients V838~Mon and V1309~Sco \citep{munari2002a, tylenda2011a}. Also, the Great Eruption of $\eta$~Carinae in the 1840s that created the famous Homunculus nebula was possibly due to the merger of two massive stars \citep{smith2018a, hirai2021a}. Merged stars can appear as blue straggler stars (BSSs) in star clusters \citep{mccrea1964a, hills1976a}, and, when stars merge, magnetic fields are produced. This mechanism is thought to be at the origin of at least some of the ${\approx}\,7\%$ of OBA main-sequence (MS) stars with strong, large-scale surface magnetic fields \citep{ferrario2009a, wickramasinghe2014a, schneider2019a}. These magnetic stars may subsequently form highly magnetic WDs and NSs at the end of their lives, so-called polars and magnetars, respectively. The remarkable SN~1987A was most likely from a merged star \citep{chevalier1989a, hillebrandt1989a, podsiadlowski1990a}, and other merged stars may result in interacting and superluminous SNe \citep{vanbeveren2013a, justham2014a, schneider2025a}. Some could even give rise to ultra-long gamma-ray bursts \citep{tsuna2025a}. Mergers of NSs and red supergiants may form the enigmatic Thorne--\.Zytkow objects \citep{thorne1975a, thorne1977a} that so far elude discovery \citep{ogrady2020a}. Merged stars can further form very massive BHs, even populating the pair-instability SN mass gap \citep{dicarlo2019a, renzo2020a}. Last but not least, the most massive stars in star clusters are likely merger products \citep{schneider2014a}, and repeated and runaway collisions in sufficiently dense clusters may create supermassive stars that could collapse into intermediate-mass BHs \citep{portegieszwart2002a, mapelli2016a, gieles2018a, vergara2025a}.

As is evident from these examples, the field of stellar mergers is vast, and not all aspects can be covered in this review. For example, we neither cover mergers involving compact objects (WDs, NSs, and BHs) nor classical CE evolution when a giant-like star engulfs a point-like companion. The latter has recently been reviewed by \citet{ivanova2013a} and \citet{ropke2023a}. Moreover, \citet{schneider2025a} give an introduction into stellar mergers and CE evolution, and delineate between these two related and in certain respects similar dynamical phases. Instead, we here focus on mergers of relatively unevolved stars before they climb the red (super)giant branch, and also lean more towards massive stars that explode in SNe, but also discuss lower-mass stellar mergers. 

The review is organised as follows: First, we discuss why and how stars merge (\refsec{sec:paths-to-merger}), before explaining the physical processes occurring in mergers (\refsec{sec:physics-of-mergers}). We then look into how merged stars continue to evolve and shed light on their ultimate fates (\refsec{sec:evolution-fates-mergers}). Finally, we review observations of the merger process and the resulting merged stars (\refsec{sec:observations-mergers}), and provide some computationally-inexpensive, approximate tools to model the outcome of stellar mergers (\refsec{sec:approximate-merger-methods}).

%
% Paths to stellar mergers in a nutshell
%
\section{PATHS\ TO\ STELLAR\ MERGERS}\label{sec:paths-to-merger}

Multiple ways can lead to stellar mergers. In this section, we explain the mechanisms leading to stellar mergers by isolated binary star evolution (\refsec{sec:mergers-from-binary-evolution}), collisions in dense stellar systems (\refsec{sec:mergers-from-collisions}), and other processes such as in triple stars (\refsec{sec:mergers-from-other}).

\subsection{Isolated binary star evolution}\label{sec:mergers-from-binary-evolution}
\begin{figure}
    \centering
    \includegraphics[width=\linewidth]{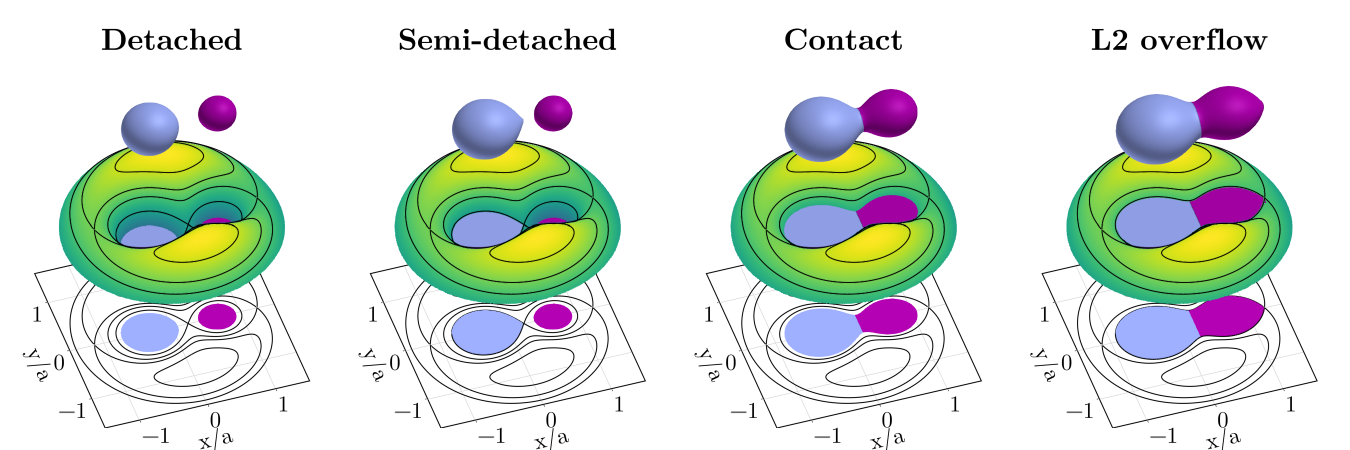}
    \caption{Illustration of the Roche potential and equipotential lines of a detached, semi-detached, contact and L$_2$ outflow binary system for a mass ratio of $q=0.5$ in the orbital $x$--$y$ plane (with $a$ being the orbital separation). The bluish star is more massive. Figure adopted from \citet{marchant2025a}.}
    \label{fig:from-rlof-to-contact}
\end{figure}

\begin{marginnote}[]
\entry{Roche potential}{The Roche potential is the combination of the gravitational and centrifugal potential of two co-rotating, static point particles. Its relative geometry depends only on the mass ratio $q$ of the two stars (\reffig{fig:from-rlof-to-contact}). When one star fills its Roche lobe (semi-detached), mass is transferred onto the companion via the inner Lagrange point L$_1$.}
\end{marginnote}
\noindent
When stars evolve, they can expand to (super)giant sizes of hundreds to thousands of solar radii. In binary star systems, the initially more massive star evolves first, and its growth can lead to a situation in which its outer layers are gravitationally attracted more strongly by the companion than by itself. The outer layers are then transferred towards the companion, and a phase of mass transfer (MT) via Roche-lobe overflow begins. Such mass transfer can proceed stably in a semi-detached binary, but it can also lead to contact phases and, in some cases, to mass loss through the outer L$_2$ and L$_3$ Lagrange points (\reffig{fig:from-rlof-to-contact}). Not all contact binaries end up in stellar mergers or CE phases. For reviews and introductory texts on binary star evolution, the reader is referred to, \eg \citet{langer2012a}, \citet{demarco2017a}, \citet{tauris2023a}, \citet{marchant2024a}, \citet{chen2024a} and \citet{marchant2025a}.

Whether a contact binary forms depends on various factors. The mass transfer (MT) case, which describes the evolutionary state of the donor star, is a crucial factor. Throughout this review, we assume that contact binaries formed by Case-A and early Case-B MT may lead to binary mergers, while those formed in late Case-B and Case-C MT rather evolve through classical CE evolution. A CE phase may itself end up in the merger of both stars. The onset of contact in massive binary stars has been studied, \eg by \citet{pols1994a}, \citet{wellstein2001a}, \citet{nelson2001a}, \citet{demink2007a}, \citet{claeys2011a}, \citet{mennekens2017a}, and \citet{henneco2024a}.

\subsubsection*{Stable mass transfer.} 
During stable mass transfer, stars can get into contact in essentially two ways. Firstly, the accretor may also expand and overfill its Roche lobe. This can either happen because of the nuclear timescale evolution of the accretor or by thermal-timescale expansion of the accretor if it cannot radiate away the gravitational energy gained from mass accretion \citep[\eg][]{kippenhahn1977a, lau2024a}. We refer to these cases as contact by `accretor expansion'. Secondly, in non-conservative mass transfer, the non-accreted mass may not be immediately lost from the binary system but could drive it into contact (\eg via L$_2$ outflow). We refer to these as binaries that cannot eject the excess matter.
\begin{marginnote}[]
Mass transfer (MT) cases:
\entry{Case A}{MT from core-hydrogen burning (MS) donors.}
\entry{Case B}{MT from post-MS donors \emph{before} core-helium ignition. Case~B MT is further subdivided into \emph{early} (B-e) and \emph{late} (B-l) for donors with a radiative and convective envelope, respectively.}
\entry{Case C}{MT from post-MS donors \emph{after} core-helium ignition}
\end{marginnote}
\begin{textbox}
\section{The three main timescales in stellar evolution}
The \emph{dynamical timescale}, $\tau_\mathrm{dyn}$, refers to the free-fall or sound-crossing timescale in stars,
\begin{equation}
        \tau_\mathrm{dyn} \approx \sqrt{\frac{R^3}{GM}} = \sqrt{\frac{3}{4\pi G \left<\rho\right>}} \approx \frac{R}{c_\mathrm{S}} \approx 0.5\,\mathrm{h}\, \left( \frac{R}{\rsun} \right)^{3/2} \left( \frac{\msun}{M} \right)^{1/2},
        \label{eq:tau-dyn}
\end{equation}
where $M$, $R$, $\left<\rho\right>$, and $c_\mathrm{S}$ are the mass, radius, mean density, and average sound speed of a star. 

The \emph{thermal} or \emph{Kelvin--Helmholtz timescale}, $\tau_\mathrm{KH}$, describes the time it takes a star to radiate away its entire internal energy, $E_\mathrm{int}$,
\begin{equation}
        \tau_\mathrm{KH} \approx \frac{E_\mathrm{int}}{L} \approx \frac{GM^2}{2RL} \approx 15\times10^6\,\mathrm{yr}\, \left( \frac{M}{\msun} \right)^{2} \left( \frac{\rsun}{R} \right) \left( \frac{\lsun}{L} \right),
        \label{eq:tau-kh}
\end{equation}
where $L$ is the luminosity of a star. 

The \emph{nuclear timescale}, $\tau_\mathrm{nuc}$, denotes the time it takes for a star to consume all its nuclear fuel, which for core hydrogen burning is
\begin{equation}
        \tau_\mathrm{nuc} \approx \frac{\Phi f_\mathrm{nuc} M c^2}{L} \approx 10^{10}\,\mathrm{yr}\, \left( \frac{M}{\msun} \right) \left( \frac{\lsun}{L} \right).
        \label{eq:tau-nuc}
\end{equation}
In this equation, $f_\mathrm{nuc}$ is the fraction of the total mass $M$ of a star available as nuclear fuel, and $\Phi$ is the mass-to-energy conversion efficiency of nuclear burning. In most cases, $\tau_\mathrm{dyn} \ll \tau_\mathrm{KH} \ll \tau_\mathrm{nuc}$. The various phases of single and binary star evolution proceed on these main timescales. For example, binary mass transfer typically proceeds on the thermal and nuclear timescales, while stellar mergers and CE evolution are dynamic phases happening on the dynamical timescale.
\end{textbox}
\subsubsection*{Dynamically unstable mass transfer.} 
Mass transfer may become dynamically unstable if the radius of the donor star expands more rapidly or shrinks less slowly than its Roche-lobe radius upon fast (adiabatic) mass loss \citep[see \eg][]{hjellming1987a, soberman1997a}. In such a case, the donor would significantly overfill its Roche lobe, which typically increases the mass transfer rate exponentially, leading to even more overfilling of the Roche lobe and runaway MT that ends up in a contact binary. Runaway MT is typically discussed for the onset of classical CE phases from late Case-B and Case-C MT, but it can also cause contact phases and induce binary mergers in Case-A and early Case-B MT.

\subsubsection*{Darwin instability.} 
Tides are important in binary evolution \citep[for reviews, see \eg][]{zahn1977a, ogilvie2014a, mathis2019a, barker2025a} and can induce contact phases and lead to stellar mergers. If the orbital angular momentum, $L_\mathrm{orb}$, is less than three times the spin angular momenta of the two stars, \ie $L_\mathrm{orb} < 3 (S_1 + S_2)$, an unstable configuration---the Darwin instability---is reached. In this case, tidal forces keep transferring orbital to spin angular momentum while being unable to reach synchronous rotation of the stars with the orbit \citep[\eg][]{darwin1879a, counselman1973a, hut1980a}. A catastrophic orbital decay sets in on a tidal synchronisation timescale. 
\vspace{\baselineskip}

\begin{figure}
    \centering
    \includegraphics[width=\linewidth]{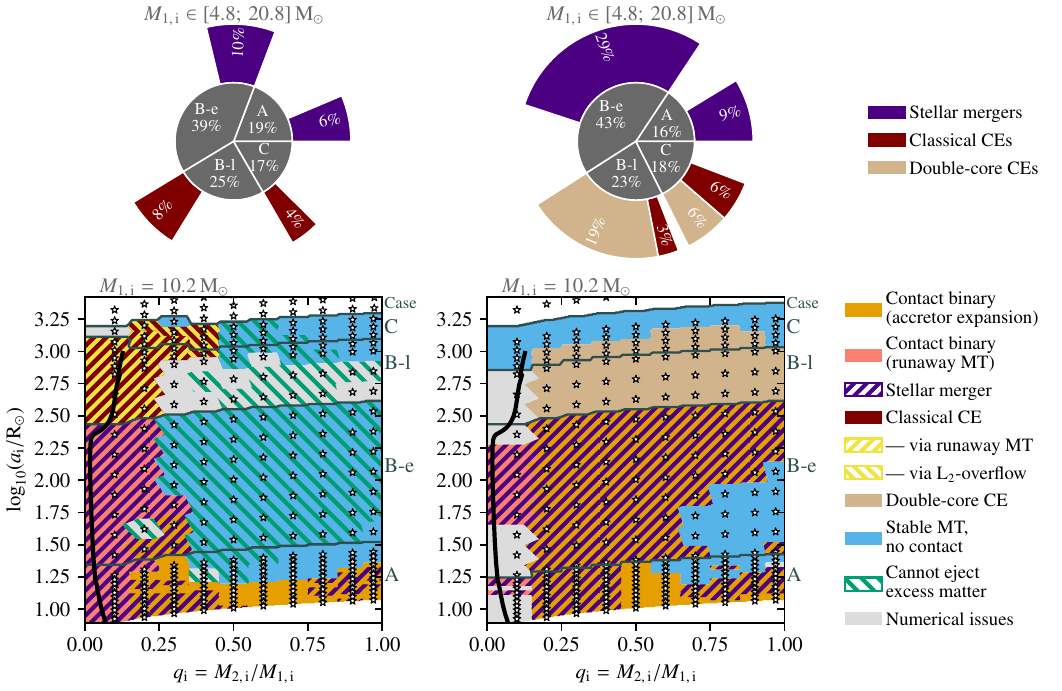}
    \caption{Occurrence of contact phases, stellar mergers and CE evolution in isolated binary stars. The bottom panels are exemplary slices of the binary parameter space in the initial mass ratio $q_\mathrm{i}$--initial orbital separation $a_\mathrm{i}$ plane for binaries with initial primary star masses of $M_{1,\mathrm{i}}=10\,\msun$. The pie charts at the top show population-averaged incidences of the main mass transfer (MT) Cases~A, B-e, B-l, and~C, and stellar mergers and classical CE events (averaged over binaries with initial primary star masses in the range $4.8\text{--}20.8\,\msun$). The left panels are for binary models using rotationally-limited mass transfer that often leads to highly non-conservative mass transfer, while the right panels are for fully conservative mass transfer (non-rotating stars). Models to the left of the solid black lines ($q_\mathrm{i}\lesssim0.1$) are Darwin unstable. For the meaning of the legend, see the main text. The models and data are from \citet{henneco2024a} and \citet{henneco2025a}. Figure courtesy of Jan Henneco.}
    \label{fig:merger-ce-incidence}
\end{figure}

\noindent
In all of these cases, mass may be lost through the outer Lagrange points L$_2$ and L$_3$, which leads to orbital shrinkage and ultimately precedes a stellar merger. The aforementioned mechanisms leading to contact binary systems are depicted in \reffig{fig:merger-ce-incidence} for primary stars of initial mass $M_{1,\mathrm{i}}=10\,\msun$ for the solar-metallicity binary models of \citet{henneco2024a} and \citet{henneco2025a}. The indicated stellar mergers are for the first contact phase. 

The two binary grids differ in their model for mass transfer. In the first set, rotationally-limited mass transfer is used, which results in non-conservative mass transfer once the accretor has reached break-up rotation at the surface \citep{langer2003a}. In these models, contact binaries from the expansion of the accretor during stable Case~A mass transfer are found only for the initially tightest binaries and from dynamically unstable, runaway mass transfer for $q_\mathrm{i}\lesssim0.25$ across all mass-transfer cases. The Darwin instability only plays a role for the smallest initial mass ratios of $q_\mathrm{i}\lesssim0.1$. For almost all of Case-B and~C mass transfer, there is not enough photon energy to immediately eject the non-accreted matter \citep[\cf][]{marchant2017a}. Mass transfer is on the thermal timescale of the donor stars and can thus be rapid enough (${\gtrsim}\,10^4\,\msun\,\mathrm{yr}^{-1}$) to fill the accretor's Roche lobe and lead to significant L$_2$ outflow \citep[][]{lu2023a, scherbak2025a}. This in turn will shrink the orbit and could drive the binary system into contact or render mass transfer unstable \citep[\cf][]{willcox2023a, picco2024a}.

The second binary grid assumes fully conservative mass transfer, and many more systems now get into contact because of the expansion of the accretor, while the binary parameter space for stable mass transfer without contact narrows down to some Case-A and~B-e systems with $q_\mathrm{i}\gtrsim0.5$. The accretors are brought out of thermal equilibrium and expand into contact configurations. The possible CE phases from late Case-B and Case-C mass transfer are denoted as double-core common-envelope evolution following \citet{ropke2023a}. It remains uncertain whether contact binaries from this thermal-timescale expansion of the accretor indeed become dynamic and then lead to stellar mergers and CE phases.

Based on these models, \citet{henneco2024a} estimate that at least 16\% and 12\% of mass-transferring binaries with $5\text{--}20\,\msun$ primaries merge and evolve through a CE phase, respectively. These numbers increase to at least 38\% and 34\% if binary mass transfer is fully conservative \citep[\reffig{fig:merger-ce-incidence};][]{henneco2025a}. Indeed, binary mass transfer might be more conservative than assumed in the rotationally-limited model \citep[\eg][]{lechien2025a}. These numbers are broadly consistent with earlier estimates \citep{sana2012a, schneider2015a}. Integrating over a uniform past star-formation rate, it is found that about 10\% of the present-day massive-star population are merger products \citep[\eg][]{podsiadlowski1992a, demink2014a}.

\subsection{Collisions in dense stellar systems}\label{sec:mergers-from-collisions}

Stars in dense environments, such as star clusters, can collide. The cross section of this process is given by the geometrical cross section, $\pi d_\mathrm{coll}^2$ (with $d_\mathrm{coll}=R_1+R_2$ being the sum of the radii of both stars), and a second term accounting for gravitational focusing,
\begin{equation}
        \sigma_\mathrm{coll} = \pi d_\mathrm{coll}^2 \left[ 1 + \frac{2G(M_1+M_2)}{d_\mathrm{coll} v_\mathrm{rel}^2} \right].
        \label{eq:coll-cross-section}
\end{equation}
In this equation, $v_\mathrm{rel}$ is the relative velocity of the two stars of masses $M_1$ and $M_2$. Gravitational focusing typically dominates in encounters where the relative velocity is smaller than the escape velocity from the stars. This is, for example, the case in most globular clusters but not necessarily in the denser nuclear star clusters. Integrating over a Maxwellian relative velocity distribution with velocity dispersion $\sigma_v$ in the limit that gravitational focusing dominates, one finds for the time between collisions of stars with mass $M_*$, radius $R_*$, and number density $n$, \citep{binney2008a}
\begin{equation}
        T_\mathrm{coll} = \frac{1}{n \sigma_\mathrm{coll} v_\mathrm{rel}} \approx 7\times10^{14}\,\mathrm{yr}\, \left(\frac{R_*}{\rsun}\right)^{-1} \left(\frac{M_*}{\msun}\right)^{-1} \left(\frac{n}{\mathrm{pc}^{-3}}\right)^{-1} \left(\frac{\sigma_v}{\kms}\right).
        \label{eq:collision-time}
\end{equation}
Hence, for collisions to occur over the lifetime of stars, the stellar density $n$ needs to be high enough; hence, collisions are expected to occur mostly in such environments.

Binary stars and higher order multiples offer a greater interaction cross section, and a subset of such interactions may either directly lead to a collision or a delayed binary merger by hardening binaries and inducing eccentricity \citep[\eg][]{heggie1975a, sigurdsson1993a, fregeau2004a, antognini2016a}. In stellar environments with a high primordial binary fraction and low stellar density, binary--binary and binary--single star interactions are more frequent than single--single encounters \citep[see \eg][]{heggie1975a, bacon1996a, davies2015b}. In some cases, runaway mergers are also possible \citep[\eg][]{portegieszwart1999a, portegieszwart2002a}.

\subsection{Other mechanisms}\label{sec:mergers-from-other}

There are further mechanisms that can lead to stellar mergers. Triple systems and other multiples are common \citep{moe2017a, offner2023a}. In hierarchical triple stars, an inner binary can exchange angular momentum with the outer tertiary companion, leading to periodic oscillations in the eccentricity of the inner binary and the inclination of the outer perturber. These are the so-called von Zeipel--Lidov--Kozai oscillations \citep{vonzeipel1910a, lidov1962a, kozai1962a} with periods much longer than that of the inner binary's orbit. In combination with tides, the eccentricity oscillations in triple stars can drive the inner binary into contact and mergers \citep[][]{perets2009a, naoz2014a, toonen2022a}. The physics of such mergers is expected to be similar to that of isolated binary mergers. Triple systems can also become chaotic, possibly leading to mergers that would be more akin to stellar collisions. For reviews on these mechanisms, see, \eg \citet{naoz2016a} and \citet{perets2025a}.

Binary stars moving through circumstellar material are subject to tidal torques that can harden the binary and ultimately lead to mergers \citep[][]{stahler2010a, korntreff2012a}. Such processes may be most relevant during star formation or in other environments with high circumstellar densities (\eg discs of active galactic nuclei). Similarly, tidal torques and resonant interactions with a circumbinary disc, \eg left by previous binary mass exchange, can harden an inner binary and induce eccentricity, ultimately leading to stellar mergers \citep{kashi2011a, tuna2023a, wei2024a, valli2024a}. This process depends strongly on the mass and lifetime of the disc.

%
% Physics in stellar mergers
%
\section{PHYSICS\ OF\ THE\ MERGER\ PROCESS}\label{sec:physics-of-mergers}

As described above, stars can merge in different ways, and the exact process triggering the merger will, at least in parts, determine its outcome. In the following, we focus on mergers from evolution in isolated binary stars---called \emph{binary mergers} from hereon---and head-on collisions of stars in dense stellar systems---called \emph{collisions of stars}. The main differences between these two types of mergers are the merger dynamics and the available energy and angular momentum. Hence, the merger process and its outcome may differ between these two merger types. In this section, we first explain the four main phases of stellar mergers before reviewing the most important physical processes occurring in them and highlighting common features as well as differences between binary mergers and stellar collisions.

\subsection{The four phases of stellar mergers}\label{sec:phases-merger}

The merger process can be grouped into four phases: (i) a contact phase leading to the final coalescence of both stars, (ii) the coalescence itself, (iii) the immediate merger aftermath that is often dictated by the co-evolution of a central merged star and a surrounding disc-like structure, and (iv) the longer-term, thermal relaxation of the merged star.

\subsubsection{From contact to coalescence}

In binary mergers, a contact phase precedes the final merger event. The duration of such phases can be long-lived, even up to the nuclear timescales of stars, as is evident from observations of long-lived low-mass W~Ursae~Majoris (W~UMa) stars \citep[\eg][]{lucy1979a, li2004a, rucinski2007a, gazeas2021a} and more massive contact binaries \citep[\eg][]{almeida2015a, abdul-masih2021a, abdul-masih2024a, vrancken2024a}. While these systems may ultimately all merge, not all contact phases end up in a stellar merger \citep[\eg in chemically homogenously evolving binary stars, see, \eg][]{mandel2016b, marchant2016a}. To facilitate a merger, angular momentum has to be extracted from the binary orbit and this happens mainly in two ways\footnote{In mergers of compact objects or stars with convective envelopes, gravitational wave radiation and magnetic braking contribute to/dominate this process.}: (i) by mass and the associated angular momentum loss, and (ii) by tides transferring orbital to spin angular momentum. 

Mass loss in deep contact configurations occurs via the outer Lagrange points L$_2$ and L$_3$. For simplicity, we only consider L$_2$ outflow that has a specific angular momentum of $j_{L_2}=r_{L_2}^2\sqrt{G M a}$, where $M=M_1+M_2$ is the total binary mass, $a$ is the orbital separation, and $r_{L_2}$ is the distance of the L$_2$ point from the centre of mass in units of $a$ \citep[$r_{L_2}\approx1.2$ within ${\approx}\,10\%$ for a large range of mass ratios $q$, see \eg][]{pribulla1998a}. A merger will occur once an orbital angular momentum of $\Delta L_\mathrm{orb} = L_\mathrm{orb}(a_{L_2}) - L_\mathrm{orb}(R_1)$ is lost; here, $a_{L_2}$ is the orbital separation when L$_2$ outflows starts, and the dynamic merger is assumed to occur latest once the orbital separation equals the radius, $R_1$, of star 1\footnote{Assuming the radius of star~1 does not change. If it does, it may stabilise the system.}. For a circular orbit with $L_\mathrm{orb}=\mu_\mathrm{red} \sqrt{GMa}$, where $\mu_\mathrm{red}=M_1 M_2/M$ is the reduced mass, a dynamical merger phase starts once a mass, $\Delta M_{L_2}$, of about
\begin{equation}
        \frac{\Delta M_{L_2}}{M_1+M_2} \approx r_{L_2}^{-2} \frac{q}{(1+q)^2} \left[1 - \left( \frac{R_1}{a_{L_2}} \right)^{1/2} \right]
        \label{eq:L2-mass-loss-to-merger}
\end{equation}
is lost through the L$_2$ point \citep[][]{macleod2017b}. For $r_{L_2}\approx 1.2$ and $R_1/a_{L_2}\approx0.5$, this amounts to 2--5\% of the total binary mass for $0.1 \leq q \leq 0.9$ that would be lost in a slow\footnote{Velocities of ${<}\,25\%$ of the binary's escape velocity \citep[][]{shu1979a, pejcha2016a, pejcha2016b}.} equatorial outflow before the final merger. Direct 3D simulations of runaway mass transfer transitioning into a dynamic CE phase confirm this picture \citep[\eg][]{macleod2018b, macleod2020a, macleod2020b}. Similarly, slow winds from contact binaries or non-accreted mass from Roche-lobe overflow can remove angular momentum at a magnitude similar to that of L$_2$ outflow, and can thus initiate the dynamical merger phase upon sufficient mass loss \citep[\eg][]{sawada1984a, scherbak2025a}. The mass lost through the outer Lagrange points is initially bound; depending on the binary properties and the exact fluid flow through the outer Lagrange points, the material either gets unbound via tidal acceleration by the central binary or stays bound to the binary, possibly forming a circumbinary disc \citep[see \eg][]{kuiper1941a, shu1979a, pejcha2014a, macleod2018a, hubova2019a}.

In the case of the Darwin instability, one also expects mass outflow through the outer Lagrange points before the final coalescence, but less mass will be lost as the orbital decay is also driven by orbital angular momentum loss by tides \citep[\cf][]{macleod2017b, macleod2020a}. In some cases, tides may even induce large-scale oscillations and deformations of the stars that could be observable \cite[\eg `polygram' stars,][]{macleod2019a}.

A catastrophic orbital decay with a subsequent merger after a preceding contact phase has been observed in V1309~Sco \citep[][]{tylenda2011a}. The merger in V1309~Sco led to the prototype of a class of transients called `luminous red novae' that will be further discussed in \refsec{sec:lrn}. For further information on the stability of close and contact binaries, and the transition to dynamical merger events, the reader is referred to the studies of, for example, \citet{lai1993c, lai1993a}, \citet{rasio1995a}, \citet{dsouza2006a}, \citet{lombardi2011a}, and \citet{hwang2015a}.

Contrarily, in stellar collisions, stars collide at their closest (periastron) passage, and they can be on elliptic, parabolic or hyperbolic orbits. Collisions are typically characterised by their impact parameter, $b=r_\mathrm{p}/(R_1 + R_2)$, that is, the ratio of the closest (pericenter) distance, $r_\mathrm{p}$, and the sum of their radii, $R_1$ and $R_2$. The stars may not immediately merge into one star at the first contact, but a first grazing encounter can be followed by others before the final merger. Hence, no long-lived contact phase precedes stellar collisions. 

The angular momentum of collisions depends on the impact parameter, and, for example, there is no angular momentum in head-on collisions ($b=0$). Elliptic orbits with semi major axis $a$ and eccentricty $e$ have an angular momentum of $L_\mathrm{orb}=\mu_\mathrm{red} \sqrt{G M a (1 - e^2)}$. Circular orbits encountered typically in binary mergers thus have an orbital angular momentum smaller by a factor of $\sqrt{1+e}$ than stellar collisions of elliptic orbits with a periastron separation $a_\mathrm{peri}=a_\mathrm{coll}(1-e)$ equal to the orbital separation $a$ of a circular binary merger. Moreover, the kinetic energy of collisions is larger than in binary mergers. For collisions on parabolic orbits with zero total energy, the kinetic energy is equal to the gravitational potential energy at each point in the orbit, and hyperbolic orbits can have arbitrarily large collision energies and are typically parametrised by the initial relative velocities of the colliding stars when they are furthest away from each other.

\subsubsection{Coalescence}

Once contact phases become dynamical, the coalescence proceeds within a few orbital periods \citep[see \eg][]{lai1993c, lai1993a, rasio1995a, dsouza2006a, lombardi2011a, hwang2015a, macleod2018b, schneider2019a, ryu2025a}. This holds both for binary mergers and collisions. The final coalescence of two stars into one object often proceeds via (partial) tidal disruption(s) of one or both stars, and gives rise to turbulent fluid flows and chemical mixing of the two stars. We will describe this process in more detail below and also provide an intuitive way to understand the mixing and the resulting structure of the merged star.

\subsubsection{Dynamical and viscous relaxation: co-evolution of merged stars with remnant disc}

Because of the usually large amount of orbital angular momentum, a disc-like structure forms around a central merged star while the merger structure relaxes dynamically. This structure can be quite massive ($\mathcal{O}(10\%)$ of the total mass), and it depends greatly on the type of merger and the available angular momentum as explained above. The disc structure likely evolves on its turbulent-viscous timescale, which is longer than the dynamical relaxation right after coalescence. For a thin disc and an $\alpha$-viscosity model \citep[][]{pringle1981a, shakura1973a}, this timescale is about the accretion timescale of the disc \citep[see also][]{shen2012a, schneider2020a},
\begin{equation}
        \tau_\mathrm{visc} \approx \tau_\mathrm{acc} = \frac{1}{3} \frac{R^2}{\alpha H^2 \Omega} \approx 0.2\,\mathrm{yr}\,\left( \frac{10^{-2}}{\alpha} \right) \left( \frac{R/H}{2} \right)^2 \left( \frac{r_\mathrm{disc}}{10\,\rsun} \right)^{3/2} \left( \frac{M_*}{\msun} \right)^{-1/2},
        \label{eq:tau-acc} 
\end{equation}
where $\alpha$ is the effective turbulent viscosity parameter, $R/H$ the ratio of the disc's radius and scale height, $r_\mathrm{disc}$ the approximate disc (radius) size, and $M_*$ the mass of the central merged star enclosed by the disc. From this estimate, we see that the central remnant may quickly accrete most of the disc material. We discuss this accretion phase and its importance for the final outcome of stellar mergers further in \refsec{sec:star-disc-evolution}.

\subsubsection{Thermal relaxation of merged star}

Parts of the orbital energy will thermalise in stellar mergers and heat the merger products such that they are initially out of thermal equilibrium. The merged stars will regain full equilibrium after about a thermal timescale (see \refeq{eq:tau-kh}). This process takes much longer than accreting the disc-like structure left behind after the dynamical coalescence and hence allows us to consider these two phases separately from each other. We come back to this last phase of a stellar merger in \refsec{sec:thermal-relaxation}.

\subsection{Merger simulations and the emerging picture of the coalescence process}

\begin{marginnote}[]
\entry{Blue straggler stars}{In coeval stellar populations, BSSs populate the region above the MS turn-off in the HR diagram---a region forbidden by classical single star evolution. Today, we know that blue stragglers are rejuvenated stars that gained mass either by mass transfer in binary systems or stellar mergers.}
\end{marginnote}

\begin{figure}
    \centering
    \includegraphics[width=\linewidth]{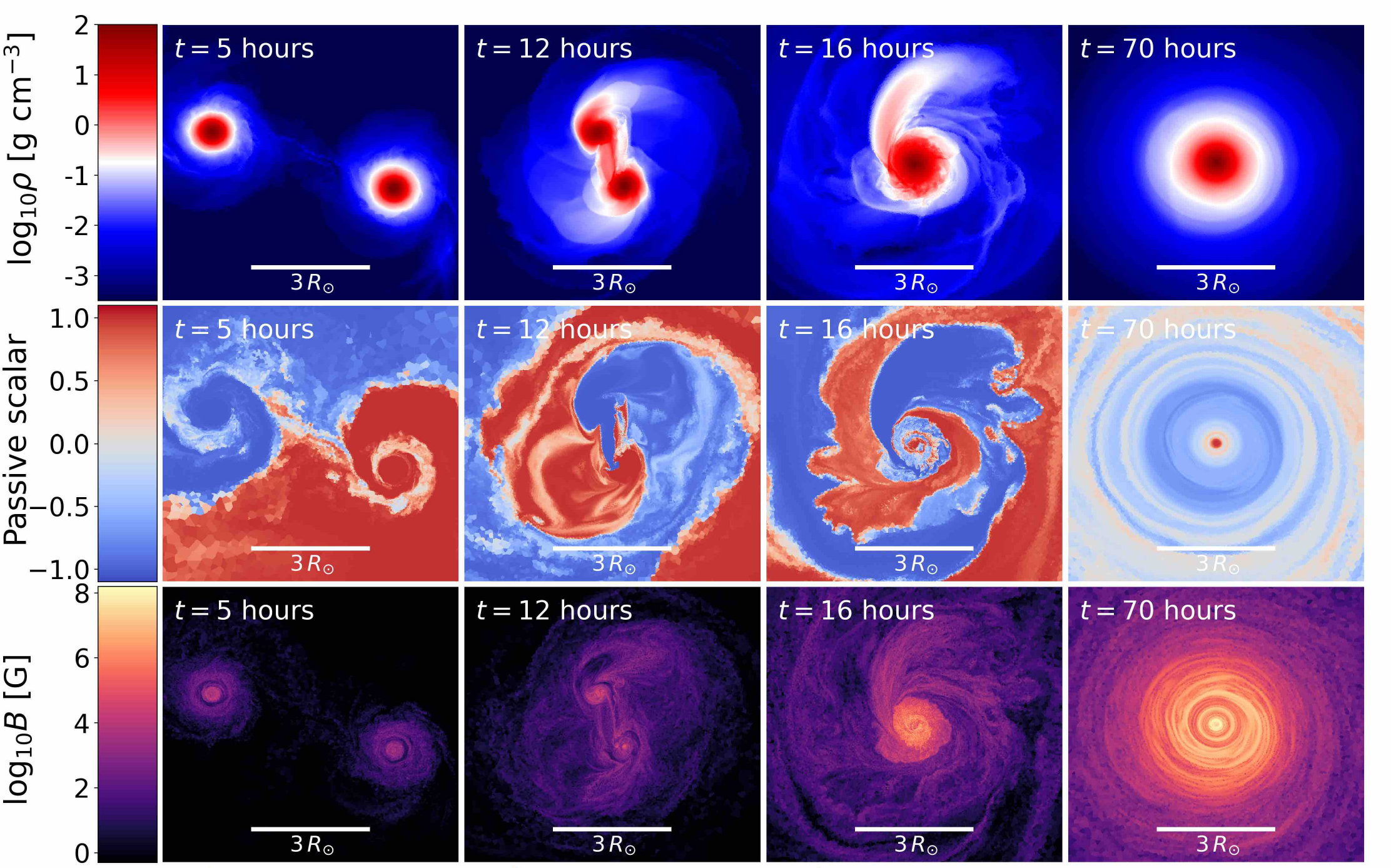}
    \caption{Stellar collision between a $0.7\,\msun$ and a $0.6\,\msun$ MS star with impact parameter $b=0.5$ from \citet{ryu2025a}. Shown are the density (upper panels), a passive scalar (middle panels) and the absolute magnetic field strength (bottom panels) at four different times $t$. Time $t=0$ corresponds to the first contact. In total, the stars experience three contact phases at $t=0$, $\approx11$ and $\approx14\,\mathrm{h}$, and they merge in their last one. Time $t=70\,\mathrm{h}$ shows the dynamically relaxed merger product. The passive scalar traces material originally from the two stars (the $0.7\,\msun$ star in red and the $0.6\,\msun$ star in blue) and thereby shows how both stars are mixed into the merger remnant. Figure adopted from \citet{ryu2025a}.}
    \label{fig:low-mass-collision}
\end{figure}

\begin{figure}
    \centering
    \includegraphics[width=\linewidth]{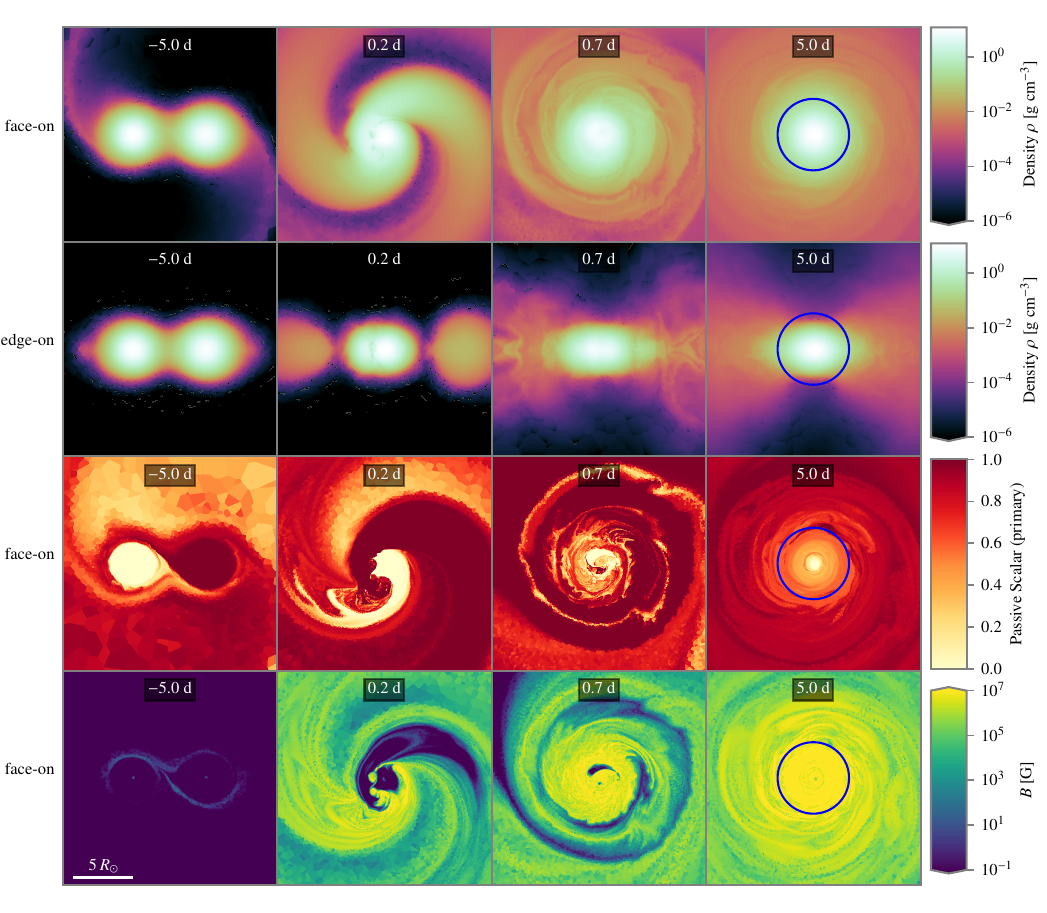}
    \caption{Similar to \reffig{fig:low-mass-collision} but for the binary merger of a $9\,\msun$ and a $8\,\msun$ MS star from \citet{schneider2019a}. The density is shown in the orbital (face-on) plane and perpendicular to it (edge-on) in the two top panels. The passive scalar tracing the matter of the more massive $9\,\msun$ star (dark red) is shown in the orbital plane in the middle panels, and the bottom panel shows the absolute magnetic field strength again in the orbital plane. Here, time $t=0\,\mathrm{d}$ is the time when both stars coalesce into one object. At $t=5\,\mathrm{d}$, the blue circle of radius $3\,\rsun$ indicates the approximate size of the central, spherically symmetric $\approx14\,\msun$ merged star that is surrounded by a $\approx3\,\msun$ disc-like structure. Figure reproduced from \citet{ohlmann2025a}.}
    \label{fig:high-mass-merger}
\end{figure}

As mentioned in the introduction, the interest in mergers of stars arose from the realisation that collisions between stars must be frequent in dense stellar systems such as globular star clusters, and that the collision products may be at the origin of the back-then enigmatic BSSs. Hence, there is a rich literature on simulations of collisions of low-mass stars, typically employing 3D SPH computations \citep[\eg][]{benz1987a, lombardi1995a, lombardi1996a, sills1997a, sills2001a, sills2002a, sills2005a, freitag2005a, ryu2024a, ryu2025a}. Also collision simulations of massive stars have been considered \citep[\eg][]{lai1993b, freitag2005a, suzuki2007a, gaburov2008a, glebbeek2013a, ballone2023a} as well as triple star collisions \citep[\eg][]{lombardi2003a, gaburov2010a}. The gained knowledge about collisions has been used to develop simplified methods that can rapidly predict the outcome of the computationally-expensive 3D simulations \citep[\eg][]{lombardi2002a, gaburov2008a, heller2025a}. Binary mergers have also been explored by direct 3D simulations using SPH and Eulerian (M)HD codes \citep[\eg][]{rasio1995a, dsouza2006a, lombardi2011a, nandez2014a, motl2017a, schneider2019a, chatzopoulos2020a, shiber2024a} but to a somewhat lesser extent than stellar collisions. In some cases, only certain phases of the merger are simulated, \eg the onset of a core merger inside a CE event \citep{ivanova2002a} or the initial phase of a merger \citep{wu2020a}. In the following, we explain the physics of stellar mergers using the examples of a low-mass, parabolic collision from \citet[][\reffig{fig:low-mass-collision}]{ryu2025a} and a high-mass binary merger from \citet[][\reffig{fig:high-mass-merger}]{schneider2019a}. Both simulations employ the moving-mesh MHD code \arepo \citep{springel2010a, pakmor2011a, pakmor2016a, springel2019a, weinberger2020a}. Stellar mergers are complex, and the outcome is difficult to predict. To obtain an intuitive picture of this process, Archimedes's principle and the idea of entropy-sorting are useful (see textbox).
\begin{textbox}
\section{Simplified picture of stellar mergers: Archimedes' principle or the concept of entropy-sorting}\label{sec:entropy-sorting}
While stellar mergers are highly complex, one can use the concept of buoyancy or Archimedes' principle to obtain some intuition and understanding of stellar mergers and their outcome. We imagine the two stars being composed of fluid elements of different densities, temperatures and chemical compositions. Each of these fluid elements has a different buoyancy, and those with the largest buoyancy will float to the surface of the merged star while those with the smallest buoyancy sink to the centre. In stellar interiors, the entropy of fluid elements, $s$, is a measure of buoyancy, and stars that are stable against buoyancy have an entropy profile that increases outwards, \ie $\mathrm{d}s/\mathrm{d}r>0$ \citep[\eg][]{landau1959a}. This entropy criterion for buoyantly stable stars is nothing but the Schwarzschild criterion for stability against convection. In the case of non-homogeneous stars with chemical gradients, the entropy criterion for stability against buoyancy reads,
\begin{equation}
        \frac{\mathrm{d}s}{\mathrm{d}r} - \left(\frac{\partial s}{\partial \mu}\right)_{\rho, P}\; \frac{\mathrm{d} \mu}{\mathrm{d}r}>0,
        \label{eq:ledoux-entropy}
\end{equation}
where $\mu$, $\rho$ and $P$ are the mean molecular weight, density and pressure, respectively \citep{bisnovatyi-kogan2001a}. This is the Ledoux criterion for stability against convection expressed in the form of entropy.

In the simplest picture, one can get an idea of the structure of merged stars by considering the entropy of the fluid elements of the two stars before the merger and assuming that the entropy does not change in the merging process. The lowest entropy fluids would then form the core of the merged star, and the highest entropy material would end up at the surface. This concept of `entropy sorting' is, of course, incomplete, and the entropies are not conserved. The merger process is not adiabatic and, for example, shocks can generate entropy, radiative losses remove heat and thus entropy, and chemical mixing also modifies the fluid's entropies \citep[see also][]{maeder2009a}. In the past, 3D merger simulations have been used to understand how the buoyancy (entropy) is affected by the merging process to then apply corrections to the pre-merger buoyancies before constructing merger structures following the idea of entropy-sorting \citep[\eg][]{lombardi1996a,lombardi2002a, gaburov2008a, heller2025a}.
\end{textbox}

As described above, collisions with non-zero impact parameter do not immediately lead to the coalescence of both stars, but there can be multiple (grazing) encounters before the two stars finally form one merged object. These encounters and the final merger structure of the collision of a $0.7\,\msun$ and a $0.6\,\msun$ MS star with an impact parameter of $b=0.5$ are shown in \reffig{fig:low-mass-collision}. Upon each encounter, some mass is ejected and energy dissipated such that the orbit tightens until both stars coalesce into one merger remnant in one of the next pericenter encounters (here, the third encounter). The coalescence itself proceeds by disrupting both stars and mixing them into a final merger remnant. In this example, the initially less massive $0.6\,\msun$ star is mixed into the more massive star, and the core of the more massive star is also the core of the merged star. The mixing of both stars is best seen by the passive scalar in the middle panel of \reffig{fig:low-mass-collision}. At each encounter, magnetic-field amplification takes place, and, in the end, the entire merged star is highly magnetised (\refsec{sec:b-field-amplification}). While not visible in \reffig{fig:low-mass-collision}, the dynamically-relaxed merged star at $t=70\,\mathrm{h}$ has an oblate shape and is not spherically symmetric because of the initial orbital angular momentum in off-axis collisions with non-zero impact parameter. This shape may suggest that a disc-like structure could form that can be crucial for the further evolution of the merged star \citep[see also][]{benz1987a, sills2005a}.

In stellar collisions with a small or even zero impact parameter, both stars may coalesce into one object upon their first encounter \citep[see \eg][]{benz1987a, sills1997a, sills2001a, freitag2005a}. For fixed total energy, the lower the impact parameter, the less orbital angular momentum is present, the more mass is ejected dynamically, the less oblate are the structures of the merged stars, and the weaker is the magnetic field amplification \citep[\eg][]{freitag2005a, ryu2025a}. Moreover, the collision dynamics depend on the collision energy, which can in principle be arbitrarily high, leading to significant dynamic mass loss (\refsec{sec:mass-loss}).

In contrast to stellar collisions, there is a longer contact phase in binary mergers. As described in \refsec{sec:paths-to-merger}, the binary gets into deeper and deeper contact before the two stars finally merge into one object. As in stellar collisions, the merger proceeds by disrupting both stars and mixing them into the merged star. \reffig{fig:high-mass-merger} shows such a merger for the $9\,\msun$ and $8\,\msun$ MS binary merger from \citet{schneider2019a}. The coalescence has similarities with tidal disruption events, and the fluid flow is highly turbulent, leading to mixing and fast magnetic field amplification. In the shown binary merger, the core of the merged star consists of the core of the initially less massive $8\,\msun$ star, a result not anticipated by the concept of entropy sorting (see \refsec{sec:entropy-sorting}) but observed also in stellar collisions with almost equal-mass companions \citep[see \eg][]{glebbeek2013a}. Most strinkingly, a massive disc-like structure ($\approx3\,\msun$) forms around the $\approx14\,\msun$ central, spherically-symmetric merger remnant as shown at time $t=5\,\mathrm{d}$ in the edge-on view of the matter density in \reffig{fig:high-mass-merger}. Such structures may be absent in stellar collisions and depend on the initial orbital angular momentum of the merger. We describe the role of such structures further in \refsec{sec:star-disc-evolution}. 

The initial conditions for simulations of binary mergers are still uncertain because of the long-lived contact phase and the binary evolution that leads to the contact configurations. So far, this has not been explored much, and improvements in the understanding of the evolution of contact phases are particularly relevant \citep[see \eg][]{fabry2022a, fabry2023a, fabry2024a}. The initial conditions are less uncertain for stellar collisions.

\subsection{Mass loss in stellar mergers}\label{sec:mass-loss}

Each of the four phases of stellar mergers can lead to mass loss. We have already discussed the mass loss associated with the contact phase of binary mergers that is predominantly in the orbital plane (see \refeq{eq:L2-mass-loss-to-merger}) and the dynamical mass loss of grazing encounters in stellar collisions that is typically also associated with the orbital plane. Next, we discuss the dynamical mass loss because of the final coalescence of two stars into one merger product, before we also consider mass loss from the last two merger phases.

As explained above, ejecta masses from head-on collisions and binary mergers can differ greatly. Depending on the exact collision energy, a large fraction of the total mass of colliding MS binaries could be ejected \citep[several 10\% and up to 100\%, see \eg][]{freitag2005a} while such values can only be reached in binary mergers with evolved stars, such as in classical CE phases. Moreover, in direct head-on collisions ($b=0$), mass loss may be largely perpendicular to the orbital plane, while, in the final coalescence of binary mergers and collisions with larger impact parameters, it can be in all directions. 

To estimate the dynamical mass loss, $\Delta M_\mathrm{dyn}$, from the coalescence of both stars into one merger star, we consider an energy argument similar to that applied to classical CE events \citep[see \eg][]{ivanova2013a, ropke2023a}. We assume that a fraction $\alpha_\mathrm{m}$ of the pre-merger orbital energy, $E_\mathrm{orb}=-G M_1 M_2 / 2 a$, can be used to unbind some mass, $\Delta M_\mathrm{dyn}$, from the merged star. As in the energy formalism of classical CE events, we write the binding energy, $E_\mathrm{bind}$, of the dynamically ejected mass using a $\lambda$ parameter such that $E_\mathrm{bind}=-G (M_1 + M_2) \Delta M_\mathrm{dyn}/\lambda R$. Equating this binding energy to the pre-merger orbital energy, we have
\begin{equation}
        \Phi = \frac{\Delta M_\mathrm{dyn}}{M_1 + M_2} = \alpha_\mathrm{m} \lambda  \frac{R}{a} \frac{q}{(1+q)^2} \equiv C \frac{q}{(1+q)^2}, \quad \text{with} \quad q=M_2/M_1.
        \label{eq:dyn-ejecta-mass}
\end{equation}
The above scaling demonstrates that the ejecta fraction is higher in binaries with larger mass ratios $q$ and less tightly bound stars, \ie more evolved stars with larger $\lambda$ parameter. From head-on collisions of massive MS stars in parabolic orbits, \citet{glebbeek2013a} find that $C=0.3$ fits their data well (in these simulations, there is no mass loss from previous grazing encounters). Dividing their models into half-age MS and terminal-age MS, one finds $C\approx0.24$ and $C\approx0.33$ \citep{schneider2025a, heller2025a}, respectively, indicating the expected larger ejecta masses in more evolved stars according to \refeq{eq:dyn-ejecta-mass}. For the $9+8\,\msun$ early MS binary merger of \citet{schneider2019a}, the dynamic ejecta fraction $\Phi$ is about a factor of 10 smaller, suggesting $C\approx0.024$. The above relation cannot capture the entire physics determining the ejecta mass, and other parameterisations of $\Phi$ have been proposed \citep[\eg][]{lombardi2002a, freitag2005a}. 

In binary mergers and collisions with sufficiently large impact parameters, the dynamical ejecta may interact with the previously lost material that is predominantly in the orbital plane. The previously ejected mass (and possibly still bound material) is expected to deflect the dynamical ejecta into the polar regions, where the gas density is the lowest. This mechanism is thought to create the bipolar ejecta morphologies of stellar mergers, which we discuss further in \refsec{sec:merger-debris}.

In the third merger phase, the accretion of material from the disc-like structure formed in the coalescence of the binary system may produce magnetically-driven bipolar outflows (\refsec{sec:star-disc-evolution}). Such outflows can contribute to the overall mass loss from stellar mergers and may shape the ejecta morphologies. This phase is still highly uncertain, and further research is needed to better understand the existence and significance of such outflows.

\begin{marginnote}[]
\entry{Luminous blue variable}{These massive, blue, evolved stars are variable and suffer episodic mass loss. Some LBVs even undergo giant eruptions. The cause for their variability and mass loss is not well understood, and the Eddington limit is thought to play a role \citep[see \eg the reviews by][]{humphreys1994a, smith2017b}.}
\end{marginnote}

During the fourth phase of stellar mergers, the merged stars can be subject to further mass loss while thermally relaxing and possibly shedding angular momentum. Some orbital energy will be converted into heat in the merged star and may lead to significant expansion of the star during its post-merger relaxation phase. The star can reach high luminosities that can drive enhanced stellar winds. In the most extreme case, very massive merged stars may temporarily approach the Eddington limit and experience enhanced mass loss, eruptions and/or pulsations as is seen in luminous blue variables. Moreover, the merged star may initially rotate rapidly and could thus shed mass from its equator when approaching break-up velocity. These mechanisms are not yet well understood.

\subsection{Chemical mixing}\label{sec:merger-mixing}

As is evident from the passive scalars in \reffig{fig:low-mass-collision} and \reffig{fig:high-mass-merger}, both stars are well mixed throughout the merger remnant except for the core regions where one often finds only the core material of one of the two stars. This will be even more so in the case of a merger of a post-MS star with an MS star, where the core of the post-MS star likely forms the core of the merged star, and the rest settles around it. While there is always plenty of mixing in mergers, this does not imply that core material from one or both stars is also mixed up to the surface of the newly formed merged star. Following the idea of entropy sorting (\refsec{sec:entropy-sorting}), regions of the original stars with similar entropies likely mix while those with very different entropies do not. Hence, for many mergers, this means that the core regions mix as well as the envelopes of both stars. Indeed, binary merger simulations and stellar collisions show that nuclear burning products such as helium and nitrogen from the cores of stars are not always found at the surface of the merged stars \citep[\eg][]{lombardi1996a, sills1997a, glebbeek2008a, glebbeek2013a, schneider2019a}. Mergers of more massive stars mix core material more efficiently to the surface because such stars have larger convective cores. Similarly, mergers of more evolved MS stars also tend to mix more core material to the surface because of the chemical gradient left by the receding convective core during the MS evolution. The mixing occurring during the dynamical phase of stellar mergers is only part of the overall mixing. Convective and thermohaline mixing, as well as other processes during the thermal relaxation of the merged star, can be similarly important.

Mixing of material in and out of the core of the merged star is crucial for further evolution. In MS star mergers, this mixing sets the amount of available nuclear fuel and thus the remaining MS lifetime of the merged stars (\cf \refsec{sec:rejuvenation}). The amount of mixing in and out of the cores of MS star collisions has been studied, \eg by \citet{glebbeek2008b} and \citet{glebbeek2013a} for low-mass and high-mass stars, respectively. In post-MS star mergers, mixing of helium-rich material from the core into the envelope is critical for the question whether the merged stars can evolve into long-lived blue supergiants (BSGs) or not \citep[see \eg][]{barkat1988a, saio1988a, hillebrandt1989a, podsiadlowski1990a, podsiadlowski1992a, ivanova2002a, vanbeveren2013a, justham2014a, podsiadlowski2017a, menon2017a, urushibata2018a, menon2024a, schneider2024a}. Simulating mergers of post-MS and MS stars is computationally difficult because of the vastly different central densities, temperatures and other scales. Hence, no full simulations of such mergers with, \eg red supergiants exist, and approximate methods need to be employed. For example, \citet{ivanova2002a} consider the interaction of a mass accretion stream from a companion with the core of a giant star and thereby study the likely mixing occurring in the central region of evolved CE mergers.

\subsection{Rejuvenation}\label{sec:rejuvenation}

Rejuvenation describes the phenomenon that a merged star appears younger to us than it truly is \citep[\eg][]{hellings1983a, podsiadlowski1992a, braun1995a, dray2007a, schneider2016a}. This can generally occur when stars accrete mass, such as by a merger as discussed here. To observe rejuvenation, comparison clocks are required (\eg provided by coeval, non-rejuvenated stars in a star cluster or a companion star). Merged stars of \emph{all} evolutionary stages can, in principle, appear younger to us than other genuine single stars that were born at the same time. Classically, one considers MS stars and connects rejuvenation to blue stragglers; for further illustration purposes, we will now consider such MS stars. 

Rejuvenation consists of two contributions and applies to stars of all masses: the \emph{true rejuvenation} that is connected to replenishing the available nuclear fuel, thereby literally rewinding the internal clock of stars, and the \emph{apparent rejuvenation} that relates to the fact that more massive (post-merger) stars have shorter lifetimes and hence younger apparent ages. We define the fractional MS age, $f$, of a star of mass, $M$, such that the current age $t = f \cdot \tau_\mathrm{MS}(M)$, where $\tau_\mathrm{MS}(M)$ is the MS lifetime of the star. Because of mixing of fresh fuel into the core region, the new apparent fractional MS age $f_\mathrm{app}<f$, and the apparent age of the merged star of mass $M'>M$ right after the merger is $t_\mathrm{app}=f_\mathrm{app} \tau_\mathrm{MS}(M')$. The relative age difference or effective rejuvenation then is
\begin{equation}
        \frac{\Delta t}{t} = \frac{t - t_\mathrm{app}}{t} = 1 - \underbrace{\frac{f_\mathrm{app}}{f}}_{\text{true rej.}} \underbrace{\frac{\tau_\mathrm{MS}(M')}{\tau_\mathrm{MS}(M)}}_{\text{app.\ rej.}}.
        \label{eq:rejuvenation}
\end{equation}
This equation nicely illustrates the two components of rejuvenation. An often applied rejuvenation model for the true rejuvenation $f_\mathrm{app}/f$ is that of \citet{glebbeek2008b} and its extension to massive stars including wind mass loss by \citet{schneider2016a}. These models incorporate calibrations for chemical mixing by the merger process based on SPH head-on collisions of low- and high-mass MS stars \citep{glebbeek2008b, glebbeek2013a}. For example, in $3+3\,\msun$ and $30+30\,\msun$ half-age MS mergers ($f=0.5$), \citet{schneider2016a} find that the rejuvenation amounts to $1 - 0.82 \cdot 0.23 \approx 0.81$ and $1 - 0.60 \cdot 0.60 \approx 0.64$, respectively. It is generally more effective in more massive stars, later MS ages and more equal-mass binaries, and of the order of the MS lifetime of the more massive pre-merger star.

\subsection{Rotation}\label{sec:merger-rotation}

In most binary mergers and stellar collisions with non-zero impact parameter, there is typically more orbital angular momentum in the orbit than can be stored in the merged star\footnote{For the pre-merger orbital angular momentum, $L_\mathrm{orb}$, to be less than the spin angular momentum of the merged star, $S_\mathrm{m}$, we need $q/(1+q)^2 \sqrt{a(1-e^2)/R_\mathrm{m}} \lesssim r_\mathrm{g}^2 $. Here, we have assumed solid body rotation near break-up velocity for the merged star, \ie $S_\mathrm{m}=r_\mathrm{g}^2 M_\mathrm{m} \sqrt{G M_\mathrm{m} R_\mathrm{m}}$, and that the mass of the merged star $M_\mathrm{m}$ is the total pre-merger mass. The radius $R_\mathrm{m}$ of the merged star is probably on the order of $a$, and typical values for the (relative) radius of gyration are $r_\mathrm{g}^2\lesssim0.1$. Unless $e$ is large or the mass ratio $q$ is small, the RHS of this equation is typically larger than $q/(1+q)^2 = 0.08\ldots0.25$ for $q=0.1\ldots1.0$. Hence, only for some Darwin-unstable binary systems with quite unequal masses or very eccentric orbits with small $L_\mathrm{orb}$, can the merged star absorb all the orbital angular momentum.}. Hence, most merged stars can be assumed to initially rotate rapidly near break-up velocity, and they will be inflated because of energy injection by the merger process. This raises the interesting question of how they manage to lose some of this angular momentum to be able to contract into a hydrostatic and thermal equilibrium configuration \citep[see \eg][]{leonard1995a, sills2005a, ryu2025a}.

In 3D simulations of stellar mergers, the merger remnant quickly approaches solid-body rotation \citep[\eg][]{schneider2019a, ryu2025a, ohlmann2025a}. Interestingly, \citet{schneider2019a, schneider2020a} find that the moment of inertia of the merged star is unusually small because the star has an overdense core and extended envelope. Relative radii of gyration with $r_\mathrm{g}^2 = \text{a few}\times10^{-3}$ are found such that the merged star can only contain a few per cent of the initially available orbital angular momentum (for solid-body rotation near break-up velocity). Thus, such merged stars would evolve into slowly rotating stars once the interior structure has relaxed and restructured, and the star has again more typical values of $r_\mathrm{g}^2\approx0.1$. This raises the question of whether it is indeed possible to transport away most of the available orbital angular momentum from the merged star while it co-evolves with a remnant disc and approaches solid-body rotation in the immediate aftermath of the merger. Simulations and models of this in the context of the immediate post-merger evolution of merged white dwarfs suggest that this is possible \citep{shen2012a, schwab2012a}.

Besides angular momentum extraction from the merged star via a disc, there are further mechanisms that can remove angular momentum from merger remnants. Magnetic fields are produced in binary mergers and stellar collisions, and the merged stars contain excess energy from the merger process. Magnetic disc locking could further efficiently extract angular momentum from the merged star \citep[][]{sills2005a}. Moreover, magnetic braking can further help spin down merged stars \citep{leonard1995a, schneider2019a, schneider2020a} and the merged star will likely also be subject to enhanced winds, further increasing angular momentum loss. The star could even approach its Eddington limit and lose large amounts of mass and angular momentum. After thermal relaxation, the merged star evolves on a long timescale during which magnetic braking is particularly efficient in spinning down stars \citep[\eg][]{meynet2011a, potter2012a, schneider2020a, keszthelyi2020a, keszthelyi2022a}. While their existence in stellar mergers is yet unclear, magnetically-driven outflows could further help regulate the angular momentum of post-merger stars (see \refsec{sec:star-disc-evolution}).

In conclusion, it is not yet clear what the rotation rates of merged stars are once they have fully relaxed. Multiple lines of argument suggest that merged stars are initially rapidly rotating and that they can spin down efficiently. They may thus be mostly observed as rather slowly rotating stars. We discuss this further in \refsec{sec:observations-mergers} when we consider observations of merger candidates such as BSSs.

\subsection{Magnetic fields}\label{sec:b-field-amplification}

It has long been suspected that magnetic fields are amplified efficiently during the merger of two stars \citep{ferrario2009a, nordhaus2011a, wickramasinghe2014a}. Such merged stars may thus be at the origin of the strong, large-scale surface magnetic fields observed in about 7\% of OBA MS stars with a radiative envelope \citep[\eg][]{landstreet1992a, donati2009a, fossati2015a, scholler2017a, grunhut2017a} and their compact remnant descendants, highly magnetic white dwarfs and neutron stars called polars and magnetars, respectively. Direct MHD simulations of the mergers of massive and low-mass MS stars have demonstrated that magnetic fields are indeed amplified to high enough values that can explain observations \citep[][see \reffig{fig:low-mass-collision} and \reffig{fig:high-mass-merger}]{schneider2019a, ryu2025a}.

The magnetic field amplification process in mergers of young stars shows many similarities with that observed in CE phases \citep{ohlmann2016b, ondratschek2022a, vetter2024a} and mergers of binary white dwarfs \citep{ji2013a, zhu2015a, pakmor2024a}, binary neutron stars \citep{price2006a, siegel2014a, ruiz2016a, kiuchi2018a, kiuchi2024a, mosta2020a, aguilera-miret2025a}, and neutron stars and white dwarfs \citep{moran-fraile2024a}. Efficient magnetic field amplification is thus a natural outcome of such dynamical events.

\begin{figure}
    \centering
    \includegraphics{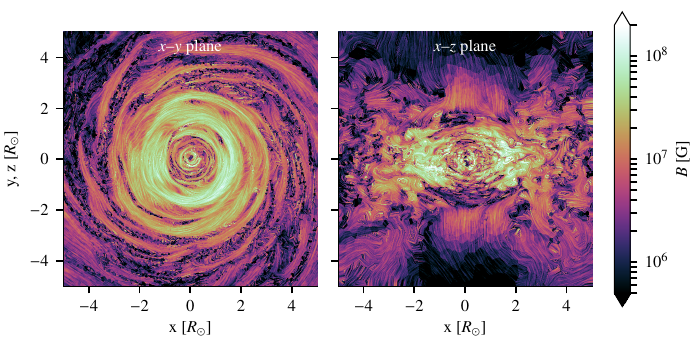}
    \caption{Absolute magnetic field strength $B$ and magnetic streamlines in the equatorial $x\text{--}y$ plane and perpendicular $x\text{--}z$ plane. The figure depicts the magnetic field $6\,\mathrm{d}$ after the final coalescence of the $9$ and $8\,\msun$ MS star merger from \citet{schneider2019a}.}
    \label{fig:b-field-streamlines}
\end{figure}

In general, the amplification mechanism is two-fold \citep[\eg][]{schneider2019a, pakmor2024a}. First, magnetic fields are amplified on small scales in shear layers by Kelvin--Helmholtz instabilities and the magneto-rotational instability \citep{balbus1991a, hawley1991a}. In binary mergers, the shear layers are formed when mass is exchanged in a deep contact binary and at the interface of the merging stellar cores when the stars are finally tidally disrupted and fully coalesce to form the new merged star. In collisions, this occurs during grazing encounters that cause shear and induce turbulent fluid motions. Second, the merged star rotates rapidly and a massive, rotationally-supported torus forms around the spherically-symmetric central merger remnant in binary mergers and in collisions with sufficiently large impact parameters. Differential rotation and fluid instabilities such as the magneto-rotational instability support an effective, large-scale $\alpha\Omega$ dynamo that further amplifies the magnetic field and transfers magnetic energy from small to large scales \citep[\ie an inverse cascade;][]{schneider2019a, pakmor2024a, ohlmann2025a}. The magnetic field strength and topology of the merger of a $9$ and $8\,\msun$ MS star from \citet{schneider2019a} is shown in \reffig{fig:b-field-streamlines}, and the ordered large-scale magnetic fields are visible (including a large-scale polar component seen in the edge-on view).

Magnetic field amplification saturates when the magnetic energy approaches energy equilibrium with the energy reservoir feeding it (\ie turbulent energy in the first amplification step and differential rotation energy in the second step). An approximate magnetic field strength $B$ for such an amplification process may thus be obtained by equating the turbulent energy density, $e_\mathrm{turb} = \rho v_\mathrm{turb}^2/2$, with the magnetic energy density, $e_\mathrm{B}=B^2/8 \pi$, \citep[\eg][]{schneider2025a}
\begin{equation}
        B \approxeq 10^3\,\mathrm{G} \, \left( \frac{M}{5\,\msun} \right) \left( \frac{3\,\rsun}{R} \right)^2.
        \label{eq:b-field-strength}
\end{equation}
Here, we have assumed that the turbulent velocity is the Keplerian rotational velocity of the merger remnant of mass $M$ and radius $R$, and that the density $\rho \propto M/R^3$. In this picture, the magnetic flux per unit mass, $\Phi_\mathrm{B}/M \propto BR^2/M$, is constant. If the so-formed magnetic OBA stars were indeed the progenitors of highly magnetic white dwarfs and neutron stars, these compact remnants should show similar $\Phi_\mathrm{B}/M$ values. Indeed, \citet{wickramasinghe2014a} was the first to note that all of these magnetic objects have similar $\Phi_\mathrm{B}/M$ values. 

The produced magnetic flux depends on the turbulent and rotational energy induced by the merger process. In stellar collisions, it thus depends on the impact parameter of the collision. Exact head-on collisions ($b=0$) are thus expected to give rise to the lowest magnetic flux, and it may even be negligible. Indeed, the collision simulations of \citet{ryu2025a} show that the magnetic energy and flux in the dynamically relaxed merger remnants are strongest for the largest impact parameters.  

It is still unclear how long these magnetic fields can persist. Dissipative processes such as magnetic reconnection or Ohmic losses may remove magnetic flux, and, right after the merger, there are indeed several magnetic loops that will likely reconnect (\reffig{fig:b-field-streamlines}). The resulting magnetic field topology from stellar mergers with a dominant toroidal component is similar to magnetic field configurations identified by \citet{braithwaite2004a} to be stable \citep[see also][for further discussions on the stability of magnetic field configurations in non-convective stars]{braithwaite2017a}. There are further open questions, such as the topology of magnetic fields after the thermal relaxation phase of the merged star, and whether strong, large-scale magnetic fields are then indeed observable and still exist.

\subsection{Star--disc co-evolution and bipolar outflows}\label{sec:star-disc-evolution}

Because of the large orbital angular momentum, a disc-like structure forms in stellar mergers. In the exemplary $9+8\,\msun$ merger of two MS stars, this structure contains about $3\,\msun$ \citep{schneider2019a}. As discussed above, the accretion of disc-like material onto the central merger remnant offers an efficient mechanism to remove angular momentum from the merged star while accreting the mass from this disc. This situation is similar to star formation through a disc where magnetic fields and magnetically-driven bipolar outflows are thought to regulate the accretion and the birth rotation of stars \citep[see \eg][]{banerjee2007a, machida2007a, carrasco-gonzalez2010a, bally2016a, hartmann2016a, kolligan2018a, oliva2023a}. Also in compact-object mergers, the star--disc co-evolution and accretion onto the central merger remnant causes magnetically-driven bipolar outflows that, in the case of binary neutron star mergers, are highly relativistic and have been suggested to power short-duration gamma ray bursts \citep[see \eg][]{siegel2014a, ruiz2016a, mosta2020a, moran-fraile2024a, kiuchi2024a, pakmor2024a}. Qualitatively similar outflows are found in MHD simulations of CE events \citep[][]{garcia-segura2018a, garcia-segura2020a, garcia-segura2021a, ondratschek2022a, vetter2024a} and are caused by a combination of the magneto-rotational Blandford--Payne mechanism \citep[][]{blandford1982a} and magnetic tower jets \citep[][]{uchida1985a, lynden-bell1994a}. Moreover, disc simulations in 2.5D around a hypothesised merger remnant show indications of such outflows \citep{moranchel-basurto2023a, moranchel-basurto2024a}. While full merger simulations with self-consistent, magnetically-driven, bipolar outflows are still lacking, one should nevertheless expect such or similar outflows to exist in the aftermath of some stellar mergers of young stars. The outflows may help regulate the final rotation rate and total mass of the merged stars.

%
% Evolution and fates of merged stars
%
\section{EVOLUTION\ AND\ ULTIMATE\ FATES\ OF\ MERGED\ STARS}\label{sec:evolution-fates-mergers}

After having described the physics of stellar mergers and the merger outcome, we now consider the further evolution. We start with the fourth phase of stellar mergers that concerns the thermal relaxation of merged stars (\refsec{sec:thermal-relaxation}), before we look into their longer-term evolution (\refsec{sec:evolution-mergers}) and ultimate fates (\refsec{sec:fates-mergers}).

\subsection{Thermal relaxation after a stellar merger}\label{sec:thermal-relaxation}

Right after the merger and after accreting most of the disc-like material (if present), the merged star is out of thermal equilibrium. Parts of the orbital energy of the former binary system have been converted into heat, which triggers an expansion of the merged star. The thermal relaxation phase depends greatly on the angular momentum content of the merged star. Too much angular momentum may prevent the merged star from contracting back into full equilibrium. Such a star would need to lose angular momentum first, \eg via a decretion disc, mass loss and/or magnetic braking \citep[\eg][]{sills2005a}. For efficient angular-momentum extraction from the merger remnant during the accretion phase of the disc-like circummerger material, this is less of an issue \citep[\eg][]{schneider2020a}. 

In all cases, the merged star will initially have an unusual interior structure: it may have a dense core region but low-density envelope, \ie an unusually small moment of inertia factor $r_\mathrm{g}^2$. Upon restructuring, \citet{schneider2020a} find that their $9+8\,\msun$ Case~A merger remnant (\cf \reffig{fig:high-mass-merger}) spins up at the surface while the star expands and then spins down during the star's contraction. The merged star might reach break-up velocity at the surface and shed additional mass. It may also evolve to so cold temperatures that the outer layers become convective. This can greatly enhance chemical mixing in the envelope and dredge up material originating from the cores of the pre-merger stars. Moreover, stars with convective envelopes can efficiently lose angular momentum, \eg by winds and spin down \citep[\cf][]{heger1998b}. During the expansion, the surface luminosity of the merged star also increases, and it may approach and possibly even exceed the Eddington limit. This could lead to enhanced mass loss and may even trigger eruptive mass loss \citep[\cf][]{owocki2019a}, similar to that seen in luminous blue variables.

\citet{schneider2020a} find that the core of the merged star is overly dense and hot compared to its full equilibrium values. Together with chemical mixing from the merger process that led to a non-equilibrium abundance of CNO elements, enhanced nuclear burning is observed during the thermal relaxation phase; this burning is in part from re-establishing CN(O) equilibrium, similar to the pre-MS evolution of stars. The enhanced nuclear burning drives a transient convective core that brings in additional fresh nuclear fuel into the core of the merged star. The magnitude of this mixing rivals that seen in the merger itself in this particular case. In mergers of post-MS stars, hydrogen that was mixed into the core could similarly lead to a transient convective core, and it may even be conceivable that some post-MS mergers turn into MS stars again by this mixing. This has not yet been observed in models.

After a few global thermal timescales (\refeq{eq:tau-kh}), the excess energy from merging has been radiated away, and the merged star has settled into a new equilibrium configuration. It might already be slowly rotating and will continue spinning down through wind mass loss and magnetic braking. The evolution after this initial thermal relaxation is described next.

\subsection{Long-term evolution}\label{sec:evolution-mergers}

\begin{figure}
    \centering
    \includegraphics[width=\linewidth]{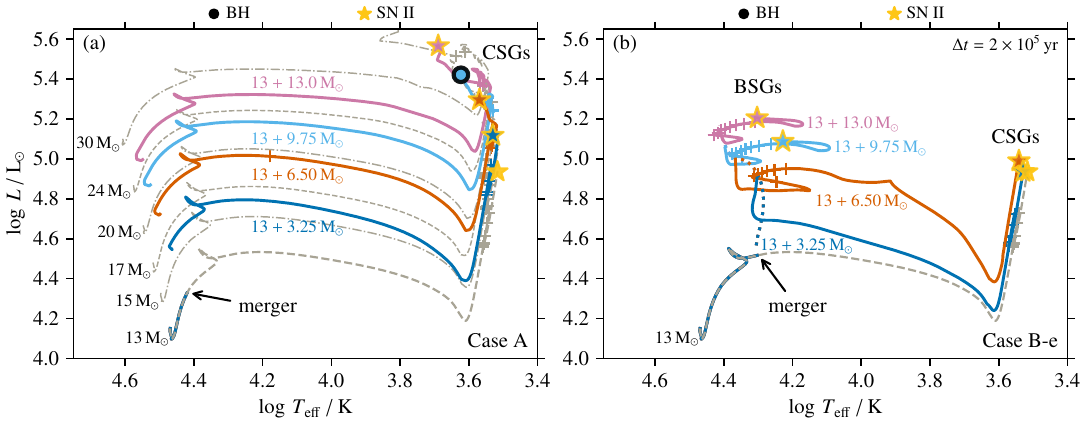}
    \caption{Hertzsprung--Russell diagram of MS+MS star (Case~A; left panel a) and post-MS+MS star mergers (Case~B-e; right panel b). In all cases, an initially $13\,\msun$ star effectively gains $3.25$, $6.50$, $9.75$ and $13.0\,\msun$ in a stellar merger. The post-merger evolution is shown until core collapse, and the likely fate of the stars is indicated (SN explosion or direct BH formation). The point in evolution when the merger happens is marked by an arrow, and plus-symbols are added to the tracks every $2\times10^5\,\mathrm{yr}$ during the post-MS evolution. In panel (a), additional single-star tracks are shown for comparison to the merger models. Models and final fate information are taken from \citet{schneider2024a}.}
    \label{fig:hrd-mergers}
\end{figure}

\begin{figure}
    \centering
    \includegraphics[width=\linewidth]{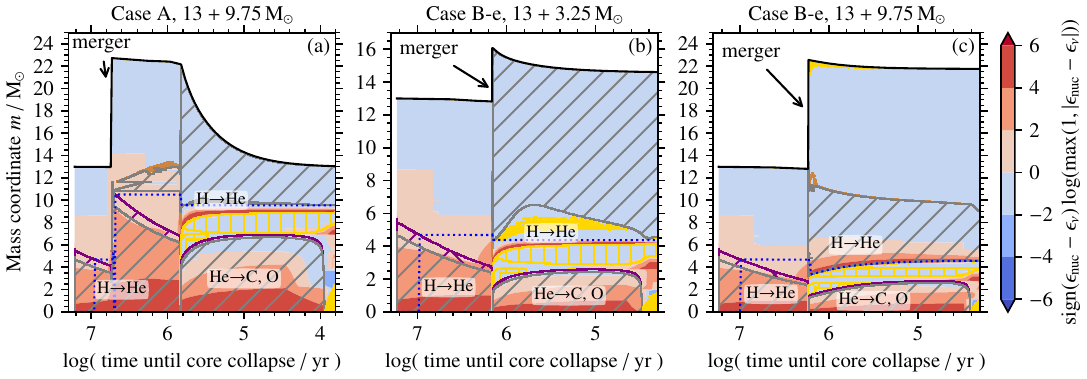}
    \caption{Kippenhahn diagrams of selected merger models from the HR diagram in \reffig{fig:hrd-mergers}. The logarithmic time axis measures time until iron core collapse to put emphasis on the core-helium burning phase of stars. The colour scale shows the specific energy gain/loss by nuclear burning, $\epsilon_\mathrm{nuc}$, and neutrino losses, $\epsilon_\nu$. The grey, purple, yellow, and brown hatched areas indicate layers of convection, convective boundary mixing, thermohaline mixing, and semi-convection. The blue dotted line denotes the helium core. The Case~A merger in the left panel (a) fully rejuvenates, adjusts its stellar structure to the new mass, and then evolves similarly to a genuine single star of the new total mass. The Case~B-e merger in the middle panel (b) burns helium in its core as a red supergiant, while the Case~B-e merger in the right panel(c) has gained enough mass to become a BSG during core helium burning thanks to convective shell-hydrogen burning.}
    \label{fig:kipp-mergers}
\end{figure}

The merger remnants of two MS stars are again MS stars, and we here focus on merged stars with a total mass ${\gtrsim}\,1.2\,\msun$ that have a convective core. These merged stars have interior structures that are reminiscent of those of genuine single stars. During the thermal relaxation, the convective core may grow into a region with a chemical gradient that is stable against convection but unstable to semi-convection. Such layers may prevent some merged stars from reaching the same convective core sizes during the MS as single stars of the same mass; they are then said to have not fully rejuvenated, and they can develop stellar structures that are not possible to achieve by single star evolution \citep[more on these below; see also][]{braun1995a}. As rejuvenation depends on the chemical gradient, not fully rejuvenating MS mergers are expected to occur more frequently in mergers with more evolved pre-merger MS stars. In most cases, it is expected that stars can almost fully rejuvenate, and the merged MS stars attain structures closely resembling those of single stars of the same mass \citep[\eg][]{glebbeek2013a, schneider2024a}. The exemplary evolution of MS merger products from primary stars of initially $13\,\msun$ is shown in the Hertzsprung--Russell (HR) diagram in \reffig{fig:hrd-mergers}, and \reffig{fig:kipp-mergers} presents a Kippenhahn diagram of the $13.0+9.75\,\msun$ Case~A merger. In this example, the merged star develops a convective core that is the same as in a single star of the same mass.

While the tracks of MS mergers in the HR diagram look similar to those of single stars of the same mass (\reffig{fig:hrd-mergers}), the merged stars will always be more luminous \citep[\eg][]{glebbeek2008b}. The mass-luminosity relation of MS stars, \eg that for homologous stars $L\propto \mu^4 M^3$, depends not only on the stellar mass, $M$, but also on the mean molecular weight $\mu$. As merged stars have, on average, a higher helium abundance and thus larger $\mu$ than single stars, they must be more luminous. Similarly, because of the lower opacity associated with the higher helium abundance in parts of their envelopes, the (super-)giant branches of merged stars are at hotter temperatures than those of corresponding single stars; also, their horizontal branches will be more luminous \citep[\eg][]{sills2009a}.

Post-MS+MS star mergers most likely lead to post-MS merger remnants, because the helium cores of the post-MS stars in the pre-merger binary have lower buoyancies and likely sink to the core of the merged stars. Some parts of the helium core will be mixed throughout the merged star, but, except for that, the helium core will probably stay intact. Hence, the merged star will have approximately a helium core mass of that of the pre-merger post-MS star, and the other parts of both stars accumulate around this core. In comparison to single stars, this almost always results in stellar structures with a smaller core-to-total mass ratio\footnote{Except for very high mass loss from the merger.}. If this core-to-total mass ratio falls below a certain threshold, the merged stars can form long-lived BSGs\footnote{A quantitative threshold for long-lived BSG remnants from stellar mergers is not known, but it will depend on the mixing of helium in the merger and the exact evolutionary state of the pre-merger binary---in more evolved systems, the threshold is at smaller core-to-total mass ratios.} \citep[\eg][]{barkat1988a, saio1988a, hillebrandt1989a, podsiadlowski1990a, podsiadlowski1992a, ivanova2002a, vanbeveren2013a, justham2014a, podsiadlowski2017a, menon2017a, urushibata2018a, menon2024a, schneider2024a}. Such merged stars may spend their entire remaining lifetimes as BSGs but can also become cool supergiants (CSGs) after an initial BSG phase, as is shown in \reffig{fig:hrd-mergers}. Not fully rejuvenated MS star mergers can form similar structures and may thus also lead to a long-lived BSG phase. \citet{menon2024a} showed that mergers can explain the observed large number of BSGs, and we discuss observations of such BSGs further in \refsec{sec:observations-mergers}.

In \reffig{fig:kipp-mergers}, we show the Kippenhahn diagrams for two post-MS+MS star mergers: in the $13.0+3.25\,\msun$ merger case, the merged star evolves into a CSG, while in the $13.0+9.75\,\msun$ merger model, the merged star remains a BSG until the star explodes in a SN (see also the HR diagram in \reffig{fig:hrd-mergers}). The Kippenhahn diagrams reveal the cause for the different evolutionary paths: whenever merged stars undergo convective shell-hydrogen burning, the stars have radiative envelopes and are compact BSGs. If shell-hydrogen burning proceeds radiatively, the star turns into a CSG with a convective envelope. \citet{schneider2024a} describe these BSG stellar structures as if an MS star with convective core-hydrogen burning is placed on top of a helium core. Indeed, these stars follow an envelope mass-luminosity relation, and convective shell-hydrogen burning is found if the temperature at the base of the hydrogen-rich envelope exceeds a certain threshold, just as in MS stars.

\reffig{fig:hrd-mergers} shows that the Case~B-e $13.0+9.75\,\msun$ merger is subject to significantly less mass loss than the equivalent Case~A merger. This is a genuine and important feature of all BSG merger remnants: the stellar wind mass losses are much lower in hydrogen-rich hot stars than in hydrogen-rich cool stars \citep[for stars not entering the LBV regime; see \eg][]{smith2014a, vink2022a}. Moreover, CSGs from mergers have larger total masses than single stars.

The stellar models shown here neglect the influence of merger-generated magnetic fields on their evolution, \eg via enhanced angular momentum transport or magnetic braking. Moreover, the magnetic fields could influence convective fluid motions inside the stars, \eg by inhibiting convective boundary mixing. It is then possible that merged stars have smaller convective cores than genuine single stars of the same mass \citep[see \eg][]{petermann2015a}.

\subsection{Ultimate fates}\label{sec:fates-mergers}

\begin{figure}
    \centering
    \includegraphics[]{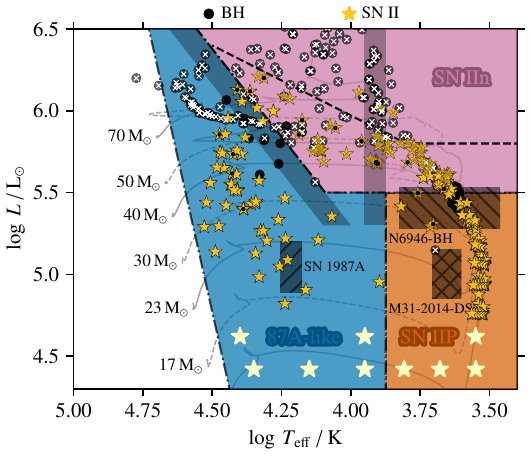}
    \caption{Hertzsprung--Russell diagram of the core-collapse positions of stellar mergers. The blue, orange and pinkish regions are an ad-hoc classification of the SNe into 87A-like, Type IIP, and interacting Type IIn SNe. Models marked by cross symbols spend more than $10^4\,\mathrm{yr}$ inside the S~Doradus instability region in the HR diagram indicated by the two light-grey shaded regions \citep{smith2004a} and above $\log\,L/\lsun=5.5$. Such models are expected to encounter enhanced mass loss and will thus likely not encounter core collapse at the indicated locations but at hotter temperatures, possibly as classical Wolf--Rayet stars. The black dashed line is the Humphreys--Davidson limit \citep{humphreys1979a}. Small black dots indicate BH formation by SN fallback. The dark grey rectangles show the positions of the progenitor of SN~1987A, and the CSGs N6946-BH and M31-2014-DS1 that may have collapsed into BHs. The slightly brighter and larger star symbols added to the bottom of the HR diagram indicate that the shown model sample is incomplete and does not properly cover the low-luminosity region. This is a modified version of the HR diagram in \citet{schneider2025a}.}
    \label{fig:hrd-sn-positions}
\end{figure}

When stars merge, their masses typically increase, directly influencing their fates as WDs, NSs or BHs. For example, some stars may grow so much in mass that they do not end as WDs anymore but evolve through all nuclear burning cycles and form an iron core that collapses when exceeding its effective Chandrasekhar mass. WDs formed from merged stars can have different masses (they may be particularly massive) and form late in comparison to those from single stars, affecting age estimates of WDs \citep[\eg][]{temmink2020a}. 

\begin{marginnote}[]
Classification of hydrogen-rich SNe used in this review.
\entry{SN IIP}{SNe characterised by a typically $\approx100\,\mathrm{d}$ long plateau in their light curves (the `P' refers to `plateau') that are linked to explosions of RSGs \citep[\eg][]{smartt2009a}.}
\entry{SN IIn}{SNe showing signs of interaction between the fast SN ejecta and slower circumstellar material. The SN light causes narrow emission lines from the slowly moving circumstellar, hence the `n' for `narrow' in SN~IIn. We use this term as a synonym and catch-all phrase for the broad class of `interacting' SNe.}
\entry{1987A-like}{SNe with a long-rising bolometric light curve and a broad light curve peak a few months after the explosion, reminiscent of the prototype SN~1987A. They are linked to explosions of BSGs.}
\end{marginnote}

For more massive merged stars, we show their final positions at iron core-collapse in the HR diagram in \reffig{fig:hrd-sn-positions}. The models of \citet{schneider2024a, schneider2025a} also include repeated mergers such that the initial masses can be tripled. The indicated outcomes of core collapse are based on the neutrino-driven SN model of \citet{muller2016a} with calibrations as in \citet{schneider2021a}. The likely resulting SN type is indicated based on the positions of the pre-SN stars in the HR diagram. All merged stars that do not undergo core-collapse as BSGs have the same pre-SN location as single stars: they follow the band of models starting on the red supergiant (RSG) branch and stretching to higher luminosities and hotter effective temperatures. Moreover, on the RSG branch, all merger models and single stars follow the same core mass-luminosity relation, while the merged stars can have much higher envelope masses. Larger envelope masses lead to prolonged plateau phases in the light curves of Type~IIP SNe \citep{popov1993a, kasen2009a, goldberg2019a}, and a merger origin might thus be a natural way to explain SNe with an unusually long plateau phase, such as in SN~2015ba \citep{dastidar2018a, schneider2025a}. \citet{temaj2024a} use the carbon-oxygen (CO) core mass, $M_\mathrm{CO}$, and find that the relation $\log (L/\lsun) = (4.372\pm0.005) + (1.268\pm0.006) \log (M_\mathrm{CO}/\msun)$ represents well all their single star models of varying convective boundary mixing parameters ($1\sigma$ statistical fit uncertainties). The very same relation also holds for the merger models in \reffig{fig:hrd-sn-positions} \citep{schneider2024a}.

In contrast, the luminosities and effective temperatures of the BSGs at core collapse are primarily given by their envelope masses and average mean molecular weight, respectively. As described above, the structures of these BSGs are reminiscent of MS stars with convective cores that are on top of a helium star, and the surface properties thus follow similar relations as for MS stars. Such merger models can naturally explain the back-then puzzling BSG progenitor of SN~1987A, its peculiar SN light curve, the asymmetric explosion, and the triple ring nebula \citep{chevalier1989a, hillebrandt1989a, podsiadlowski1990a, podsiadlowski1992c, podsiadlowski2017a, morris2007a, morris2009a, menon2017a, urushibata2018a, menon2019a}. Merged stars thus contribute greatly to the observed SN diversity \citep[for a recent work on SN light curve diversity of binary stars, see][]{eldridge2018a}.

Of all massive stars, the models burning helium in their core as BSGs and reaching core collapse as such objects have the highest final masses. Hence, if they can avoid excessive LBV-like mass loss, they form the most massive BHs or lead to SNe with the highest ejecta masses \citep[\eg][]{renzo2020c, schneider2024a}. Such stars can even form BHs populating the pair-instability SN mass gap \citep[\eg][]{dicarlo2019a, renzo2020c, costa2022a, ballone2023a}. Moreover, if these BSGs rotate sufficiently fast at core collapse, the collapsing star could accrete via a disc onto the central compact remnant, possibly powering an ultra-long gamma-ray burst \citep{tsuna2025a}.

The RSGs and CSGs entering the SN~IIn region in \reffig{fig:hrd-sn-positions} are expected to encounter enhanced mass loss. There are no RSGs observed above $\log\,L/\lsun\,{\approx}\,5.5$ in M31, the Milky Way and the Magellanic Clouds \citep{davies2018b, mcdonald2022a}, and this may be related to such stars losing mass and evolving to hotter effective temperatures. While most of the models marked by cross symbols in \reffig{fig:hrd-sn-positions} might not explode at the indicated position but could lose their entire hydrogen-rich envelope and then explode as a stripped-envelope SN, some stars that only entered this instability region shortly before core collapse could explode there. Such stars would be LBVs at the time of explosion with luminosities of up to $10^6\,\lsun$ (\reffig{fig:hrd-sn-positions}) and encounter enhanced mass loss; they may even show eruptions and have dense circumstellar media just before their terminal explosions. They are thus promising candidates for interacting SNe and possibly even superluminous SNe of Type~II that are powered by interaction with circumstellar material \citep{vanbeveren2013a, justham2014a, schneider2025a}. 

\begin{marginnote}[]
\entry{Compactness parameter}{The compactness parameter $\xi_m$ is the dimensionless ratio of mass $m$ in $\msun$ and radius $r(m)$ in $1000\,\mathrm{km}$ inside the core of a pre-SN star. Critically high compactness signals stellar structures that are prone to collapsing into BHs, while low $\xi_m$ values indicate successful SN explosions \citep[see \eg][]{oconnor2011a}.}
\end{marginnote}

The SN outcomes (\eg NS remnant masses, explosion energies, nickel yields and NS kick velocities) of stars are dictated by their core properties at core-collapse. The central specific entropy, iron core mass or compactness parameter $\xi_m$ are useful proxies to predict them \citep{temaj2024a, burrows2024a, schneider2025a}. Single stars and merged stars of varied evolutionary histories result in pre-SN stars with similar ranges of core properties and hence SN outcomes. To first order, MS mergers rejuvenate, evolve as single stars of the same mass and thus result in the same pre-SN structures. In post-MS mergers, the CO core masses are within 10--20\% the same as in the post-MS merger progenitor (but typically somewhat less massive). Merged stars exploding as BSGs finish core helium burning with a lower core carbon abundance than the equivalent single star and thus produce pre-SN structures that are more difficult to explode \citep{schneider2024a}. Contrarily, merged RSG SN progenitors have similar core structures to their pre-merger stars. The progenitors of SNe~IIP and 87A-like SNe cover similar ranges in these core proxies and CO core masses, and will thus lead to similar explosion properties and NS remnant masses. In contrast, the progenitors of SNe~IIn are from stars with typically more compact and more massive CO cores, and thus explode with somewhat higher energies, produce more nickel, have more massive NS remnants and faster NS kicks. For further details of the pre-SN structures of merged stars, see \citet{justham2014a} and \citet{schneider2024a, schneider2025a}.

The above discussion and works neglect that merged stars are possibly highly magnetic, maybe even up to the pre-SN stage. The latter likely depends on the interplay of convective zones inside the star and the magnetic fields left behind from the merger. For a rotating $3\,\msun$ star, \citet{quentin2018a} find that large-scale magnetic-field components concentrate in radiative zones, and this is how a merger-induced magnetic field could persist until core collapse. If it does, a strongly magnetised core aids the SN explosion and leads to more energetic explosions \citep{matsumoto2022a, varma2023a}.

Merged stars stem from originally lower mass stars with longer lifetimes. This fact together with the possibility of prolonging the lifetime of merger products by rejuvenation, implies that some merged stars can explode in SNe after all genuine single stars did so already. \citet{zapartas2017a} find that the last single stars explode $\approx50\,\myr$ after the formation of a coeval stellar population while this may extent up to $\approx200\,\myr$ thanks to merged stars and other binary products. Delayed SNe from merged stars can thus enhance the feedback to their surrounding environments and impact it on larger scales, as the nearby gas would likely have been cleared by previous SNe already.

%
% Observations of stellar mergers
%
\section{OBSERVATIONS\ OF\ STELLAR\ MERGERS}\label{sec:observations-mergers}

Merger products are not easy to identify. In many cases, the merged stars have stellar structures that hardly differ from those of genuine single stars, for example, in the case of MS mergers. A list of indicators or signs for a possible merger history is provided in the textbox \emph{Merger signs}. Regardless of these difficulties, certain classes or groups of stars have been linked to stellar mergers, and several well-studied individual stars likely have such an origin. This section is about these objects. A formal proof of a past merger is hardly possible (if at all), and all mentioned stars should be considered \emph{merger candidates}.
\begin{textbox}
\section{Merger signs}
The non-exhaustive list below contains several indicators that may signal a merger history of a star. However, not all of them apply to all merged stars, and other evolutionary histories or physical processes can lead to similar features. It is thus very challenging to reliably identify merger products. 

\paragraph*{Rejuvenation and age discrepancies:} Whenever a star accretes mass, it rejuvenates (\refsec{sec:rejuvenation}). This applies to both accretors of binary mass transfer and stellar mergers. Rejuvenation induces age discrepancies with respect to other stars that were born at the same time but did not gain mass. Hence, to find rejuvenated stars and thus merger candidates, one needs a comparison clock that provides the true age of the star. These clocks may be provided by a distant companion star or a coeval parent stellar population such as a star cluster. To further distinguish between accretors and mergers, one can assess the binary status of the star. In the case of accretors, the former donor star may still be around, while such a star cannot exist in the case of a merger. Measuring rejuvenation is probably the most reliable way to identify merger candidates. 

\paragraph*{Unusual rotation:} As is evident from this review, merged stars likely have unusual rotation rates. Right after the merger, the stars are likely rapid rotators that may then spin down efficiently and eventually become slow rotators. However, there remain large uncertainties on the expected rotation rates of merger products, and further research is needed to settle this question.

\paragraph*{Chemical surface abundances:} Because of the mixing in the merger process and during the thermal relaxation, nuclearly-processed and partially-burnt material may show up on the surface of merger products. Telltale chemical abundances include enhancements in helium and nitrogen with a simultaneous decrease in carbon and oxygen because of hydrogen-burning via the CN(O) cycle. Moreover, chemical elements produced or destroyed by the pp chains (\eg lithium) or the neon-sodium (NeNa) and magnesium-aluminium (MgAl) cycles could show up at the surface of merger products.

\paragraph*{Magnetic fields:} In the merger process, magnetic fields are efficiently amplified (\refsec{sec:b-field-amplification}). They may be visible as strong, large-scale surface magnetic fields or mainly exist in the deep interior of stars. Finding unusually strong magnetic fields in stars is thus a sign of a past merger.

\paragraph*{Circumstellar nebulae:} The ejecta of stellar mergers can form circumstellar nebulae that are visible for a relatively short duration (up to $\mathcal{O}(10^4\,\mathrm{yr})$) until they have dispersed to such a degree that they become unobservable. Some of the ejecta may also accumulate in longer-lived discs around merger products. 

\paragraph*{Asteroseismic fingerprints:} Some merger products have unique interior structures that result in characteristic asteroseismic fingerprints and that cannot be readily explained by single stars (\refsec{sec:merger-seismology}).
\end{textbox}

\subsection{Luminous red novae}\label{sec:lrn}

\begin{figure}
    \centering
    \includegraphics[width=\linewidth]{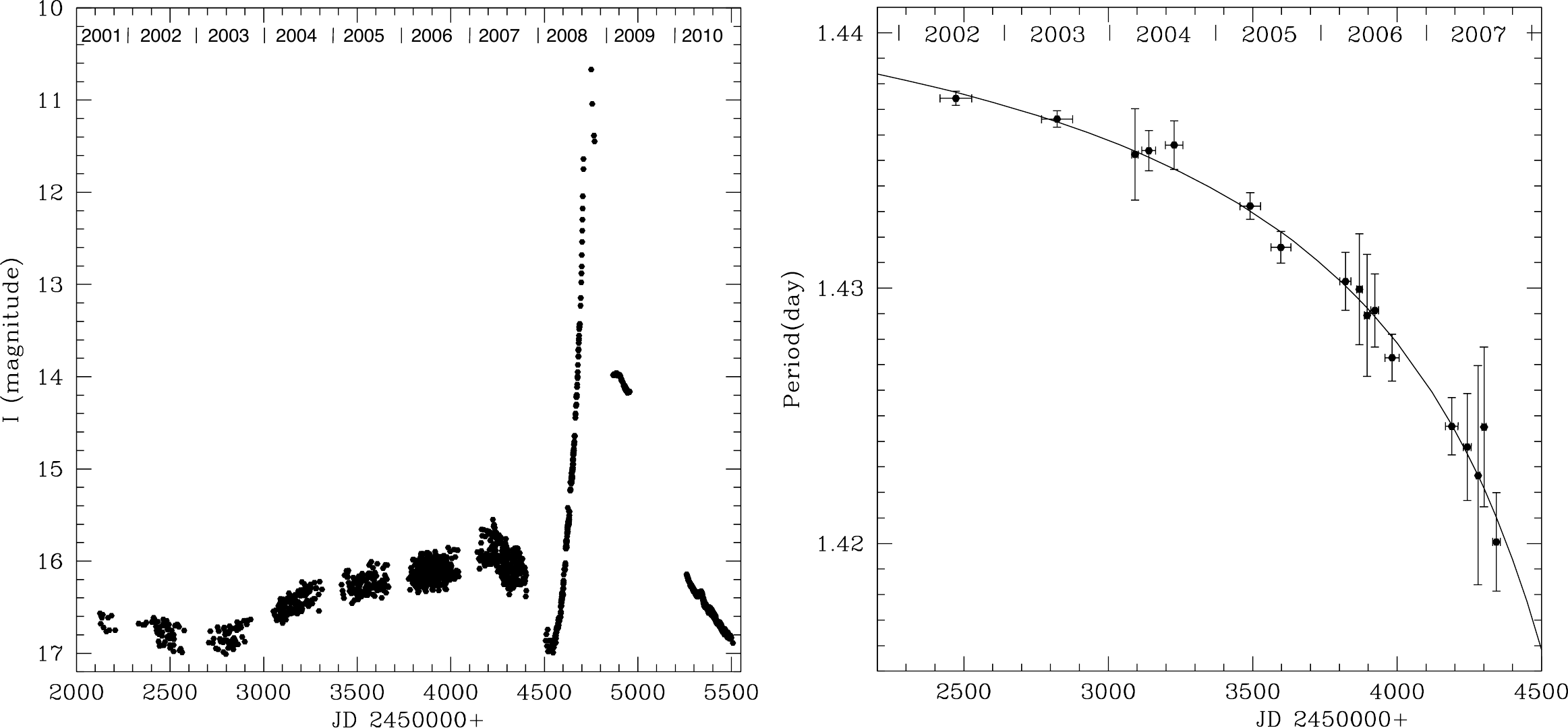}
    \caption{Optical $I$-band light curve (left) and evolution of the orbital period (right) of the LRN V1309~Sco that erupted in 2008. The solid line in the right panel is an exponential fit to the observations. Figure adopted from \citet{tylenda2011a}.}
    \label{fig:v1309-sco}
\end{figure}

When stars merge or enter CE evolution, orbital energy is released rapidly and powers an electromagnetic transient. That these transients are associated with the class of luminous red novae (LRNe)\footnote{Over the years, they received multiple names, \eg `red novae', `optical transients', `V838 Mon-type eruptions', `gap transients', `intermediate luminosity optical transients', and `SN imposters'.} became clear thanks to the prototype LRN V1309~Sco in 2008 (\reffig{fig:v1309-sco}). Using pre-outburst photometry from the Optical Gravitational Lensing Experiment (OGLE) survey, \citet{tylenda2011a} discovered the exponential decay of the orbital period of a $1.4\,\mathrm{d}$ contact binary that led to the outburst in 2008. \citet{soker2003a, soker2006a} and \citet{tylenda2006a} had already earlier suggested a merger origin of LRNe for V838~Mon that erupted in 2002 \citep{munari2002a}.

The merging contact binary in V1309~Sco had a total mass of $1\text{--}3\,\msun$ and contained a K-type primary star \citep{tylenda2011a}, while the progenitors of the two other prototypical events, V4332~Sgr and V838~Mon, most likely were $1\text{--}2\,\msun$ and $5\text{--}10\,\msun$ MS stars, respectively \citep{tylenda2005b, tylenda2005a}. More massive blue/yellow Hertzsprung-gap (HG) stars of $12\text{--}18\,\msun$ \citep{blagorodnova2021a}, ${\approx}\,18\,\msun$ \citep{blagorodnova2017a}, and ${\approx}\,30\,\msun$ \citep{smith2016b} are inferred for the progenitors of AT~2018bwo, M101-OT2015-1, and NGC-4490-OT, respectively. Intriguingly, \citet{smith2016b} note that the spectra of NGC-4490-OT and V838 Mon resemble that of the Great Eruption of $\eta$~Carinae as seen from light echoes, prompting them to speculate that it is also linked to a stellar merger. 

The V1309~Sco event has been explained by an initially low-mass binary system with primary star mass $1\text{--}2\,\msun$ becoming Darwin unstable once the initial secondary overtakes in evolution because of mass accretion and starts its ascent of the giant branch. The binary system is driven into contact by tides, eventually causing L$_2$ outflow that then led to rapid coalescence once the orbit shrank sufficiently \citep[\cf \refsec{sec:phases-merger};][]{stepien2011b, pejcha2014a, henneco2024a}. Runaway L$_2$ outflow can explain various features of the observed light curve \citep{pejcha2014a, pejcha2016a, pejcha2016b, pejcha2017a, macleod2020a}. For example, it naturally predicts the observed variations in the phased light curve if the binary is observed near edge-on\footnote{The outflowing material from L$_2$ passes through the line-of-sight to the observer once per orbit.}, and it provides a way to understand the long ${\approx}\,150\,\mathrm{d}$ rise-time of the light curve to maximum light (\reffig{fig:v1309-sco}), which is much longer than the dynamical timescale of the binary of $P_\mathrm{orb}\approx1.4\,\mathrm{d}$. The expanding L$_2$ outflow obscures the binary, and it brightens because of an expanding photosphere and the onsetting runaway mass loss. Shocks in the outflows heat the ejecta and power the initial transient. This interpretation predicts a dependence of the light curve on the line-of-sight towards the merging binary system. 

After the coalescence and further mass loss, the light curve of V1309~Sco drops from maximum light and reaches a $20\text{--}30\,\mathrm{d}$ plateau (or a second peak in other events)---a phase that \citet{ivanova2013a} model as an expanding photosphere with energy release from recombination (inspired by light curve models of Type~IIP SNe). These two phases of the light curves of LRNe explain the often found two peaks or one peak and a plateau \citep[\eg][]{matsumoto2022a}, and, in general, more luminous LRNe are associated with more massive progenitors \citep[][]{kochanek2014b, matsumoto2022a}. Dust is likely formed in LRNe and impacts the light curves and the dimming after the plateau phase \citep{pejcha2017a, macleod2022a}. \citet{hatfull2025a} compute the light curve of a merger in a CE event through direct simulations and find generally good agreement with observations of V1309~Sco.

These examples demonstrate the invaluable insights one can gain into the coalescence process of binary stars. Today, LRNe are routinely observed also outside the Milky Way \citep[\eg][]{jencson2019a, pastorello2019a}. This offers to study other individual events in great detail, such as M31-LRN-2015 in the Andromeda galaxy \citep{macleod2017b}, but also allows for population studies to, \eg assess the rates of these events \citep[\eg][]{kochanek2014b} and the progenitor connection. Moreover, observations will impose further constraints on the ejecta masses of stellar mergers \citep[\cf][]{matsumoto2022a} and allow for the investigation of bipolar outflows \citep[\cf][]{kaminski2018a}. One difficulty is to distinguish between stellar mergers of more compact stars and classical CE events involving (super)giants. \citet{macleod2022a} suggest that optical transients with smaller ejecta masses and less extinction are rather associated with mergers. In comparison, infrared transients with larger ejecta masses and high extinction are more likely CE ejections.

\subsection{Merger debris and bipolar nebulae}\label{sec:merger-debris}

\begin{figure}
    \centering
    \includegraphics[width=\linewidth]{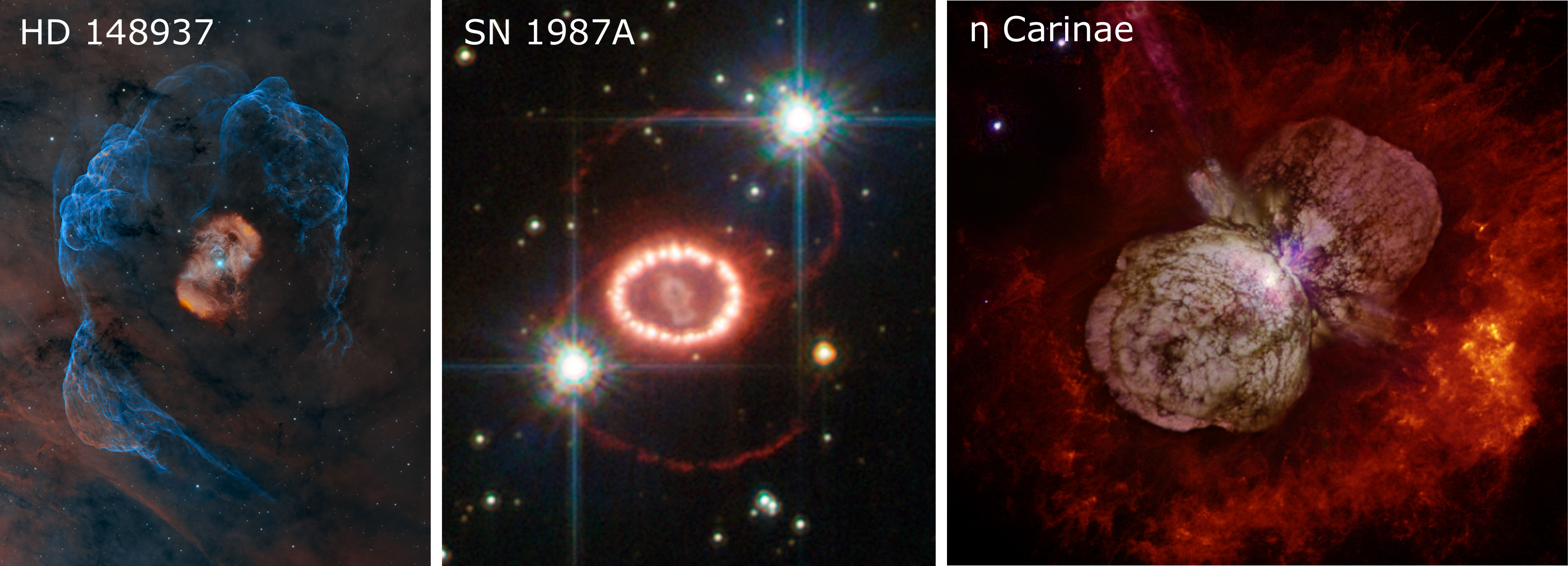}
    \caption{Exemplary bipolar nebulae surrounding the merger candidates HD~148937, SN~1987A (triple ring nebula), and $\eta$~Carinae (Homunculus nebula). Image credits and copyright: Rowan Prangley; Nathan Smith \& NASA/ESA; ESA/Hubble \& NASA.}
    \label{fig:merger-debris}
\end{figure}

As explained in \refsec{sec:mass-loss}, each merger phase can lead to mass loss that will form a nebular structure surrounding a merged star. These nebulae are expected to be bipolar as the first dense ejecta is usually in the equatorial plane, \eg via L$_2$ outflows, that then deflect subsequent ejectiles into the low-density polar region perpendicular to the orbital plane of the progenitor binary system \citep[see \eg][]{morris2007a, morris2009a, macleod2018a}. Magnetically-driven bipolar outflows and spherically symmetric enhanced winds from the central merger remnants can further sweep up the previous ejecta and shape the nebulae. In head-on collisions ($b=0$), there might not be a dense equatorial ejecta component that can deflect later merger ejecta, possibly giving rise to more spherically symmetric nebulae. \reffig{fig:merger-debris} shows three exemplary nebulae surrounding the merger candidates HD~148937, SN~1987A, and $\eta$~Carinae. 

\begin{marginnote}[]
\entry{B[e] phenomenon}{The B[e] phenomenon of forbidden optical emission lines and strong infrared excess is present in various stellar sources such as pre-MS Herbig AeBe stars, compact planetary nebulae, symbiotic stars, and B-type supergiants \citep{lamers1998a}.}
\entry{FS~CMa stars}{Since the definition of the B[e] phenomenon, several stellar sources remained unclassified (unclB[e] stars) for which \citet{miroshnichenko2007a} coined the name FS~CMa stars based on common properties linked to strong mass loss and dust formation as a consequence of binary evolution.}
\end{marginnote}

Other examples of bipolar nebulae around merger candidates include the B[e] supergiant R4 in the SMC \citep{pasquali2000a, langer2012a, wu2020a}, Sheridan~25 \citep{podsiadlowski2010c}, MN13, MN18 and MWC~349A \citep{gvaramadze2010a, gvaramadze2015a, gvaramadze2012b}, the LBV HD~168625 with an almost twin nebula of SN~1987A \citep{smith2007d}, SBW1 \citep{smith2007e}, and TYC\,2597-735-1 with its hour-glass nebula from a merger some 1000 years ago \citep{hoadley2020a}. Also, the class of FS~CMa stars, a group of lower luminosity stars showing the B[e] phenomenon, has similar nebulae and is thought to be the result of mergers \citep[\eg][]{delafuente2015a, korcakova2022a, korcakova2025a, dvorakova2024a}. For further merger candidates with bipolar nebulae, see \citet{smith2007e} and \citet{gvaramadze2015a}. Moreover, bipolar outflows have also been observed in some LRNe \citep{kaminski2018b}, and thus seem to be a characteristic feature of stellar mergers.

Common envelope phases are similar to stellar mergers and are known to produce bipolar ejecta \citep[see \eg][]{garcia-segura2018a, garcia-segura2020a, garcia-segura2021a, zou2020a, ondratschek2022a, vetter2025a}. These events may form a large fraction of the bipolar (proto) planetary nebulae \citep{soker1994a, balick2002a, demarco2009a, boffin2019a}.

\subsection{Blue stragglers, blue lurkers and their evolved descendants}

Blue straggler stars and their descendants of later evolutionary phases, the yellow and red stragglers, are thought to either form by mass accretion in binary stars or stellar mergers \citep{mccrea1964a, hills1976a}. In this review, we only consider those BSSs that form by merging; however, it is challenging to disentangle the mass transfer and merger origin observationally. For reviews on BSSs, see \citet{mathieu2025a} and \citet{wang2025b}.

The binary fraction of BSSs is purely given by binary physics and cluster dynamics, and does not contain information anymore about the primordial binary fraction of the stellar population from which they formed \citep[][]{schneider2015a}. It is essentially the number of BSSs formed by binary mass transfer divided by those formed by mass transfer and mergers. If BSSs were formed only by isolated binary evolution, their binary fractions would immediately tell us about the fraction of stellar mergers from binary evolution. In the open clusters M\,67 and NGC\,188, \citet{geller2015a} and \citet{mathieu2009a} find BSS binary fractions of $79\pm24\%$ and $76\pm19\%$, respectively, that are similar to those of field BSSs \citep{preston2000a, carney2001a}. These binary fractions are significantly higher than those found in other stars (\eg the MS turn-off stars in M\,67 and NGC\,188), and suggest that the majority formed by binary mass transfer. Indeed, \citet{geller2011a} confirm this observationally by finding the binary-stripped donor stars in orbits around the BSSs. Contrarily, the binary fraction of BSSs in the denser globular clusters are lower. \citet{wragg2024a}, \citet{muller-horn2025a} and \citet{giesers2019a} find values of $13.6\pm5.1\%$, $10.9\pm4.8\%$, and ${>}57.5\pm7.9\%$ in $\omega$~Centauri, 47~Tucanae, and NGC\,3201, respectively. Also in these cases, the binary fraction of BSSs is higher than among other cluster members ($1.2\pm0.1\%$, $2.4\pm1.0\%$, and $6.8\pm0.7\%$, respectively), showing that it is unrelated to the primordial binary fraction of the parent stellar population. The lower BSS binary fraction in globular clusters suggests that stellar collisions contribute relatively more to the BSS population than in open clusters, as expected from the higher cross section for collisions in denser environments (\cf \refeq{eq:coll-cross-section}). 

\begin{figure}
    \centering
    \includegraphics[width=\linewidth]{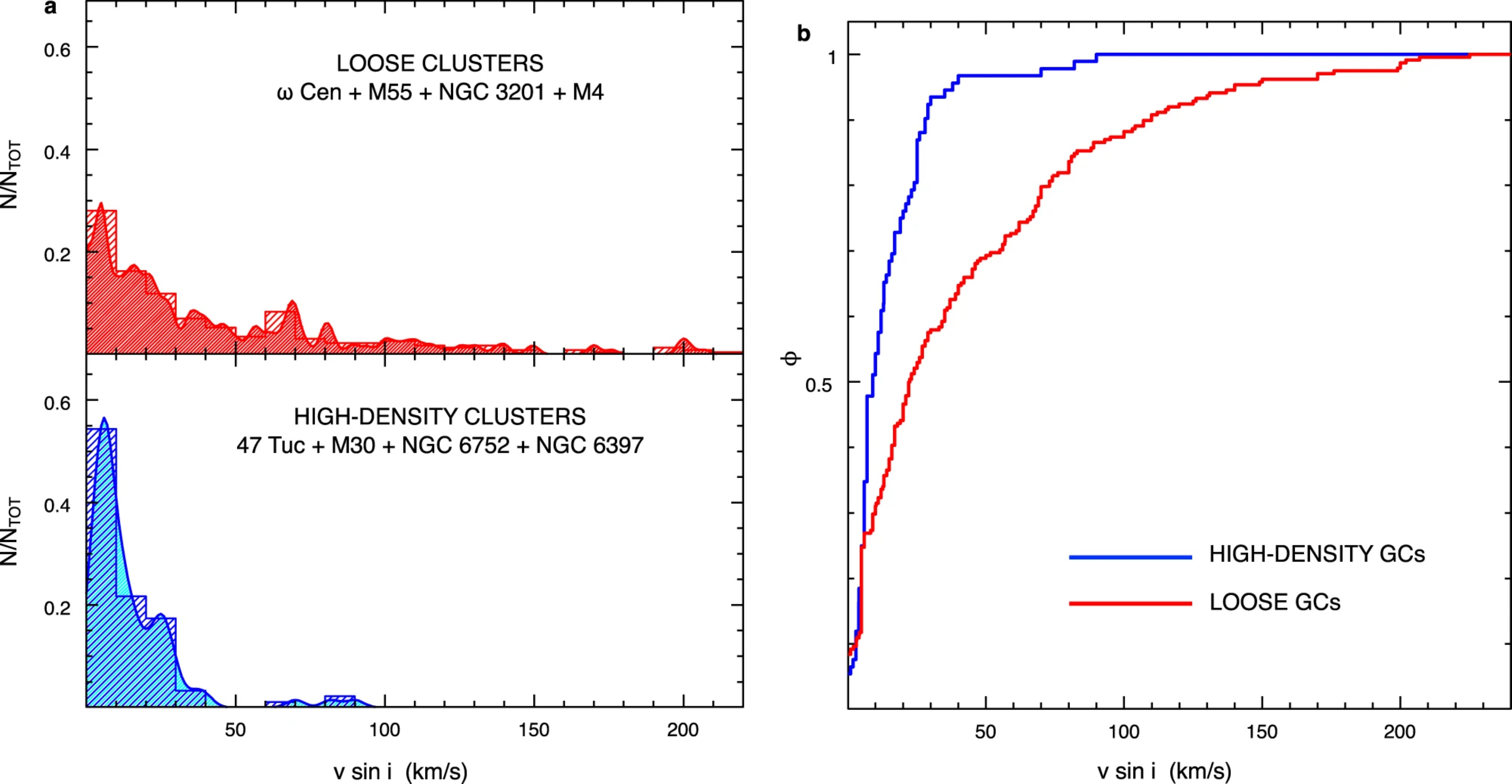}
    \caption{Distributions of projected rotational velocities, $v\sin i$, of BSSs in lower and higher-density globular clusters (GCs). Panel (a) shows the fractional number distributions, while panel (b) is for the cumulative distributions. Figure adopted from \citet{ferraro2023a}.}
    \label{fig:bss-rotation}
\end{figure}

The rotation rates of BSSs cover the entire spectrum from slow to fast \citep{lovisi2010a, lovisi2013a, lovisi2013b, bodensteiner2023a}. Blue stragglers from binary mass transfer are thought to be rather rapid rotators \citep{demink2013a}, while those from stellar mergers could be slowly rotating (\refsec{sec:merger-rotation}). Indeed, \citet{mathieu2009a} find that the BSSs in NGC\,188 that mostly formed by binary mass transfer rotate on average faster than other MS stars of the same effective temperature. Moreover, \citet{ferraro2023a} show that the BSSs in denser globular clusters with a higher fraction of collisionally formed BSSs rotate slower than in less dense clusters (\reffig{fig:bss-rotation}). Taken together, there is evidence that BSSs from stellar mergers might indeed be slow rotators. 

In the younger open cluster NGC\,330, \citet{bodensteiner2023a} show that BSSs rotate on average slower than other stars. This is particularly true for the apparently single BSSs that have the highest chances of being merger products. Mergers can also produce stars that are not more luminous than the MS turn-off in a cluster but are only bluer. Such `blue lurkers' \citep[][]{leiner2019a} stand out in split MS bands in colour magnitude diagrams of star clusters and amount to 10--30\% of the cluster stars \citep{milone2016a, milone2018a, li2017a}. \citet{wang2022a} suggest that these blue MS stars are slowly rotating merger products, which has been confirmed observationally \citep[\eg][]{kamann2023a}. 

Sub-populations of slowly-rotating A and B-type stars have been known for a while \citep[\eg][]{zorec2012a, dufton2013a, kamann2020a} and might also be merger products. In B-type stars in the Large Magellanic Cloud, \citet{hunter2008a} find that some of these slowly-rotating stars are also strongly nitrogen-enriched. Stellar mergers may explain this enigmatic group of stars that otherwise seems to defy predictions of rotational mixing in massive stars \citep[see \eg][]{brott2011b}.

As discussed in \refsec{sec:merger-mixing}, surface enrichment by stellar mergers in MS stars is more likely in more massive stars and may thus not be present in lower-mass BSSs, while it might be responsible for the nitrogen enrichment in slowly-rotating B-type stars. Indeed, only six out of 42 BSSs in 47~Tucanae show carbon and oxygen-depleted surfaces \citep{ferraro2006a}, and four out of 15 in M30 \citep[][]{lovisi2013b}. The latter four belong to the red sequence of BSSs that is thought to form from binary mass transfer. In other clusters, no depletion has been found at all \citep[][]{lovisi2013a}. Hence, there is currently no evidence for carbon and oxygen depletion in BSSs from stellar mergers.

When BSSs evolve away from the MS, they can show up as yellow and red stragglers, \ie stars that populate the region between the MS and the red giant branch. Evolved BSSs with ${\gtrsim}\,2\,\msun$ have luminosities on the red (super) giant branch that are higher than those of coeval single stars. Hence, the lowest luminosity red (super)giants in coeval stellar populations are genuine single stars, while those from mergers are more luminous---a situation similar to stars near the MS turn-off. \citet{britavskiy2019a} find that coeval single-star RSGs cover a range of about $0.2\,\mathrm{dex}$ in $\log L/\lsun$. In contrast, the range of luminosities of observed RSGs in the star clusters Hodge\,301 and SL\,639 in 30~Doradus and NGC\,2100 in the Milky Way is much larger. They attribute this to binary products such as stellar mergers. Using a luminosity-age relation, they infer the cluster ages from the least luminous RSG and show that this new technique gives results consistent with other methods.

\subsection{The most massive stars}\label{sec:most-massive-stars}

Classical BSSs are often associated with old stellar populations such as in globular clusters. However, they exist in every stellar system, including some of the youngest star clusters. For example, isolated binary evolution in a stellar population of initially $10^4\,\msun$ predicts that the most massive MS star is a merger product or accretor of binary mass transfer after about $2\pm1\,\myr$, and, after $5\,\myr$, the most massive ${\approx}\,10$ MS stars have such an origin \citep{schneider2014a}. Similarly, collisions in dense stellar systems can quickly (within a few Myr) build up massive stars. The most massive stars experience the strongest gravitational friction, sink towards the cluster centre and can thereby experience repeated collisions, a process coined runaway collisions by \citet{portegieszwart1999a}. In rich, young star clusters, such a process could even lead to the formation of supermassive (blue straggler) stars that may ultimately collapse into intermediate-mass BHs \citep[\eg][]{portegieszwart2002a, mapelli2016a, gieles2018a, vergara2025a}.

Observationally, it is difficult to identify such massive BSSs in young stellar populations because the hot BSSs are difficult to differentiate from the still hot MS turn-off based on optical photometry and the resulting colour magnitude diagrams. Moreover, the stellar populations are typically less well populated than, \eg in globular clusters. \citet{schneider2014a, schneider2015a} propose to construct present-day mass functions where the MS turn-off is indicated by a pile-up from wind mass loss in young massive stars and thereby reveal a high-mass tail beyond the pile-up consisting of unresolved multiples and likely mass gainers, stellar mergers and collision products. These features may have been identified in the Arches and Quintuplet star clusters in the centre of the Milky Way, suggesting that their 5--10 most massive MS stars could be binary products \citep{schneider2014a}. Some of these WNh stars indeed appear to be younger than less luminous cluster members \citep{martins2008a, liermann2012a}, supporting the idea that they could be rejuvenated mass gainers. Further characterisation of these very massive stars, as \eg carried out by \citet{clark2018a, clark2018b}, will eventually reveal their true nature. Similarly, some of the most massive stars in the local Universe, the initially ${>}\,200\text{--}300\,\msun$ stars in the star cluster R136 in the Large Magellanic Cloud \citep{crowther2010a, bestenlehner2020a, brands2022a}, could be the results of stellar mergers and collisions \citep[see \eg][]{portegieszwart1999a, gaburov2008b, banerjee2012a, schneider2014a}.

In all cases, the existence of merger products among the most massive stars in virtually all stellar clusters and associations can affect the determination of cluster ages when not properly accounted for. Similarly, they need to be considered when determining the initial mass function of stellar populations and a possible maximum birth mass of stars. As eluded to in \refsec{sec:fates-mergers}, the most massive MS stars may not be the most massive stars at core collapse because of post-MS mass loss (\eg because of LBV-like mass loss in stars crossing the Humphreys--Davidson limit). Hence, the most massive stars at core collapse and thus the most massive BHs formed by stars may still originate from post-MS and MS star mergers, forming long-lived BSGs that never develop unstable envelopes and enhanced mass loss.

\subsection{Blue supergiants}\label{sec:blue-supergiants}

From a single-star perspective, the blue supergiant phase is associated with stars crossing the HG on a thermal timescale. Hence, for every such HG star, there should be $\tau_\mathrm{nuc}/\tau_\mathrm{KH}\approx100\text{--}1000$ MS stars. Observationally, this expectation is not met \citep[][]{fitzpatrick1990a, castro2014a, castro2018b, deburgos2023a}, giving rise to the so-called BSG problem. Some of these BSGs may be core-hydrogen burning stars, possibly nearing the end of the MS, and others could be post-MS stars on a blue loop during core-helium burning. 

As explained in this review, mergers of post-MS and MS stars can create long-lived BSGs, and they may help solve the BSG problem. \citet{bernini-peron2023a} show that merger models can reproduce the HR-diagram positions of some observed BSGs. \citet{menon2024a} collect observations of 59 BSGs from the literature and find that about a third have helium-enriched surfaces and more than half have increased nitrogen and decreased carbon and oxygen surface abundances. The projected rotational velocities are typically ${\lesssim}\,60\,\kms$. They find that their group~3 BSGs with the highest nitrogen-to-carbon, $N/C$, and nitrogen-to-oxygen, $N/O$, number ratios can only be explained by merger models. As these stars make up more than 40\% of the BSG sample, they conclude that binary mergers are a promising way to help resolve the BSG problem. If true, these observations suggest that BSGs from mergers of post-MS and MS stars also have rather moderate rotational velocities. This finding is further corroborated by \citet{deburgos2024a}, who spectroscopically analysed an even larger sample of luminous blue Galactic stars.

If a significant fraction of the BSGs covering the HG region in the HR diagram are indeed merged stars, they should not have close binary companions. Pulsational variability makes it challenging to measure radial velocities and hence the binary fraction in BSGs \citep{simon-diaz2024a}. Taking this into account, \citet{simon-diaz2024a} find that the binary fraction of Galactic BSGs is lower by about a factor of 5 compared to non-supergiant O-type stars of similar luminosity that should represent an earlier evolutionary phase. \citet{simon-diaz2024a} suggest that a similar stark decrease in the observed binary fraction of BSGs in comparison to earlier type O stars can be found in the data presented by \citet{dunstall2015a} for 30~Doradus. If these trends can be confirmed by other studies, this could further support the interpretation of a large merger fraction among BSGs.

B[e] supergiants, sgB[e], are B-type stars showing the B[e] phenomenon (\cf \refsec{sec:merger-debris}). \citet{pasquali2000a} and \citet{podsiadlowski2006a} suggest that their circumstellar matter is from a recent stellar merger that formed the central BSG, similar to the case of the BSG progenitor of SN~1987A. These stars may thus be the immediate predecessors of the above-mentioned BSGs.

\subsection{Chemically peculiar stars}

Some chemically peculiar, low-mass stars have been suggested to be connected to stellar mergers and more broadly to binary interactions \citep[see \eg][]{demarco2017a, izzard2018b}. These include, for example the lithium-rich \citep{zhang2013a, casey2019a, singh2025a} and $\alpha$-rich young giants \citep{martig2015a, hekker2019a}. Lithium-rich giants could be related to mergers of helium WDs with red giants \citep{zhang2013a, zhang2020a} that may also produce carbon-deficient giants \citep{maben2025a}. Giants with high $[\alpha/\mathrm{Fe}]$ and young ages ($\lesssim6\,\mathrm{Gyr}$) are not predicted by standard Galactic chemical evolution models \citep[\eg][]{minchev2013a}. Hence, these stars could be evolved BSSs. As these stars pulsate, asteroseismology offers the possibility to study their interior and further our understanding of the outcome of mergers (see \refsec{sec:merger-seismology}).

\subsection{Magnetic stars}\label{sec:magnetic-stars}

As discussed in \refsec{sec:b-field-amplification}, some of the 7\% of massive OBA stars with strong, large-scale surface magnetic fields are likely merger products. This picture is further corrobated by an apparent dearth of magnetic massive stars in close binaries\footnote{A few magnetic stars are found in close binaries ($P_\mathrm{orb}\lesssim10\,\mathrm{d}$) and there is one binary, $\epsilon$~Lupii, with two magnetic components \citep[][]{shultz2015a}.} \citep{carrier2002a, alecian2015a}. Magnetic OBA stars typically rotate rather slowly, but a few rotate rapidly \citep{shultz2018a, sikora2019b}. For example, the magnetic component in Plaskett's star is a rapid rotator in a $14.4\,\mathrm{d}$ binary system and observations thereof are still puzzling to understand \citep{grunhut2022a}. Magnetic B-type stars are further connotated with surface nitrogen enhancements and boron depletion \citep{morel2008a}. Moreover, some pre-MS Herbig Ae/Be stars also have strong surface magnetic fields \citep{alecian2013a}. Many of these observational characteristics can be explained by stellar mergers at various evolutionary stages, while some observations, such as magnetic stars in very close binaries, are challenging to understand.

As mentioned above, age discrepancies caused by rejuvenation in mergers are probably the most direct way to probe the merger-origin hypothesis of (magnetic) MS stars. To this end, comparison clocks are needed. \citet{schneider2016a} tested this idea on the magnetic stars $\tau$~Scorpii and HR\,2949, finding that $\tau$~Scorpii and HR\,2949 indeed appear to be younger than the parent Upper Scorpius association and the binary companion HR\,2948, respectively. For $\tau$~Scorpii, \citet{schneider2019a} carried out 3D MHD simulations of a $9+8\,\msun$ MS star merger (\cf \refsec{sec:physics-of-mergers}) to demonstrate the viability of the merger hypothesis. Furthermore, the merger remnant is found to be broadly consistent with $\tau$~Scorpii's observed characteristics, such as the position in the HR diagram and its slow rotation. 

The discovery of a companion in a wide $25.8\,\mathrm{yr}$ orbit around the magnetic ${\approx}\,30\,\msun$ star HD\,148937 further cemented the idea that stellar mergers are responsible for at least some of the magnetic massive stars \citep{frost2024a}. The binary system is surrounded by an expanding, nitrogen-rich bipolar nebula (\reffig{fig:merger-debris}), and the magnetic component shows a clear age discrepancy with respect to its binary companion that can be understood as a result of rejuvenation in a stellar merger. This system shows almost all signs of a past merger. The kinematic age of the nebula of a few $10^3\text{--}10^4\,\mathrm{yr}$ implies that we witness a very young merger product. The magnetic star in HD\,148937 has a projected rotational velocity of $165\pm20\,\kms$. So already $10^3\text{--}10^4\,\mathrm{yr}$ after the merger, the star does not rotate rapidly anymore and is expected to further slow down via magnetic braking \citep[\eg][]{keszthelyi2019a, keszthelyi2020a}.

A strong surface magnetic field of ${\approx}\,6\,\mathrm{kG}$ has also been discovered in the FS~CMa star IRAS~17449+2320 \citep{korcakova2022a}. The circumstellar matter of these B[e] stars has been associated with the ejecta of a stellar merger (\refsec{sec:merger-debris}), providing direct evidence for merger-generated magnetic fields. Further evidence for this comes from the detection of a $43\,\mathrm{kG}$ surface magnetic field in the $2\,\msun$ quasi Wolf--Rayet star HD\,45166 with a projected rotational velocity of $\lesssim10\,\kms$ \citep{shenar2023a}. This star may be the result of a merger of two lower-mass helium stars. As discussed in \refsec{sec:b-field-amplification}, mergers of WDs generate strong magnetic fields. Some WD merger remnants are hot OB subdwarfs, and \citet{shultz2021a} and \citet{dorsch2022a} indeed find strong magnetic fields in the slowly-rotating helium-rich OB subdwarfs HD\,144941 and Gaia~DR2~5694207034772278400, respectively. All of these sources support the idea of merger-generated magnetic fields and suggest that merger remnants are rather slowly rotating stars.

If the magnetic fields of these stars persist until the end of their lives (\cf \refsec{sec:fates-mergers}), they may form highly-magnetic WDs and NSs, so-called polars and magnetars, respectively \citep{ferrario2009a, ferrario2015b, ferrario2020a, wickramasinghe2014a, schneider2019a}. Indeed, among the magnetic WDs, \citet{moss2025a} find evidence for two populations and identify the younger and more massive magnetic population with WDs that inherited their magnetic fields from, \eg merged stars.

\subsection{Merger seismology}\label{sec:merger-seismology}

\begin{textbox}
\section{Asteroseismology in a nutshell}
Many stars oscillate, and the oscillations are characterised by the three quantum numbers $n$, $l$ and $m$, which are the number of nodes in radial direction, the number of nodal lines on the surface, and the number of nodal lines that cross the equator, respectively. Oscillation modes are classified according to their restoring force: if this is predominantly the pressure gradient, we speak of pressure (p) modes, and if this is buoyancy, we call them gravity (g) modes. Other restoring forces include the Coriolis and Lorentz forces. The p and g modes can propagate in specific cavities, which are determined by the Lamb, $S_l$, and the Brunt--V{\"a}is{\"a}l{\"a} or buoyancy, $N$, frequencies, and become evanescent outside. For example, g modes can only propagate in radiative regions of a star.

For p modes with high radial orders $n$, the period spacing $\Delta \nu$ of consecutive radial orders of the same $l$ and $m$ degrees, also called the large frequency spacing, is asymptotically given by the inverse of the sound crossing time of a star,
\begin{equation}
        \Delta \nu = \left( 2 \int_0^R\,\frac{\mathrm{d}r}{c_\mathrm{s}} \right)^{-1}.
        \label{eq:delta-nu}
\end{equation}
This quantity is thus a measure of the stellar mean density (\cf \refeq{eq:tau-dyn}). In solar-like oscillators, one further finds the frequency of maximum power,
\begin{equation}
        \nu_\mathrm{max} = \frac{M/\msun}{(R/\rsun)^2 \sqrt{T_\mathrm{eff}/T_{\mathrm{eff},\odot}}} \nu_{\mathrm{max},\odot},
        \label{eq:nu-max}
\end{equation}
where the $\odot$-symbol indicates solar values. This quantity is proportional to the acoustic cut-off frequency that is related to the mass and radius of stars.

Similar to the large frequency spacing of p modes, one defines the asymptotic period spacing of g modes of consecutive radial orders,
\begin{equation}
        \Delta \Pi_l = \frac{2\pi^2}{\sqrt{l(l+1)}} \left( \int_\mathcal{R}\,\frac{N}{r} \mathrm{d}r \right)^{-1},
        \label{eq:delta-pil}
\end{equation}
which is related to the buoyancy travel time. The integral in this equation is over regions $\mathcal{R}$ in which g modes can propagate, and $\Delta \Pi_l$ typically contains information about the convective cores of stars, such as their masses. Deviations from this asymptotic value, studied by looking at so-called period spacing patterns (\ie period spacings between modes of consecutive radial order), give us information about, \eg chemical gradients and near-core rotation rates. For further information on asteroseismology, we refer to \citet{aerts2010a}, \citet{aerts2021a}, \citet{kurtz2022a}, and \citet{bowman2024b}.
\end{textbox}
Some merged stars have interior structures that are genuinely different from those of single stars. Whenever this is the case, they may have asteroseismic properties that (i) allow for their identification as merger products and (ii) enable us to study their interior in great detail and thereby learn about the outcome of stellar mergers\footnote{The same is true for non-merger, post-binary-mass-exchange stars, \eg accretors and binary-stripped stars, that have interior structures different from genuine single stars \citep[see \eg][]{deheuvels2022a, li2022a, wagg2024a, miszuda2025a}.}. As explained above, MS merger products can adjust their interior structure to the new mass and approach structures that are similar to those of genuine single stars of similar mass, albeit with a higher helium abundance. Conversely, mergers of post-MS and MS stars have structures that single stars can never achieve (\eg unusual core-to-envelope masses). Some BSSs, in particular the pulsating SX~Phoenicis stars, have been studied asteroseismically, and evidence for an increased helium abundance has been found \citep[see \eg][]{templeton2002a, cohen2012a, daszynska-daszkiewicz2025a}.

\begin{figure}
    \centering
    \includegraphics[width=\linewidth]{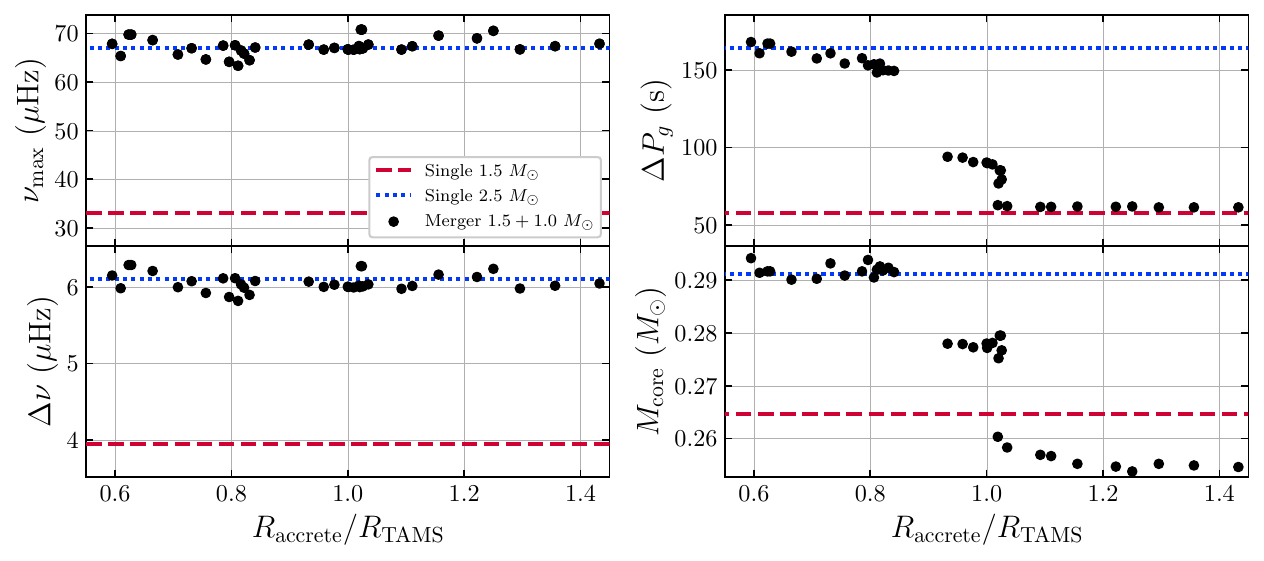}
    \caption{Frequency of maximum power, $\nu_\mathrm{max}$, large frequency spacing, $\Delta \nu$, asymptopic gravity-mode period spacing, $\Delta \Pi_l = \Delta P_\mathrm{g}$, and core mass, $M_\mathrm{core}$, of $1.5$ and $2.5\,\msun$ single stars and $1.5+1.0\,\msun$ stellar mergers on the RGB at a luminosity of $L=60\,\lsun$. For the stellar mergers, the $x$-axis shows the radius of the $1.5\,\msun$ primary star, $R_\mathrm{accrete}$, at the time of the merger in units of its terminal-age MS radius, $R_\mathrm{TAMS}$, \ie mergers of post-MS and MS stars are at $R_\mathrm{accrete}/R_\mathrm{TAMS}>1$.}
    \label{fig:rui2021a-fig3}
\end{figure}

In mergers of $1.0\text{--}1.5\,\msun$ stars, \citet{rui2021a} find that post-MS and MS star mergers have significantly lower asymptopic gravity-mode period spacings, $\Delta \Pi_l$, on the red giant branch (RGB) than genuine single stars of the same mass if the post-merger mass exceeds ${\approx}\,2\,\msun$. In such cases, the merged star develops the approximately same degenerate core of its post-MS progenitor, while the genuine single star has a more massive, non-degenerate core. As $\Delta \Pi_l$ traces the core mass of RGB stars, such stars have a surprisingly low $\Delta \Pi_l$ for a massive star (\reffig{fig:rui2021a-fig3}). If the total post-merger mass ${\lesssim}\,2\,\msun$, the interior structure is found to be similar to that of single stars and the asteroseismic properties become hard to distinguish. In all cases, \citet{rui2021a} find virtually identical frequencies of maximum power, $\nu_\mathrm{max}$, and large frequency spacings, $\Delta \nu$, from pressure mode oscillations in their merged stars on the RGB (\reffig{fig:rui2021a-fig3}). Using this knowledge, \citet{rui2021a} identify several merger candidates among RGB stars by searching for gravity-mode period spacings that are reminiscent of low-mass RGB stars in seemingly massive ($\gtrsim2\,\msun$) stars. 

In a similar vein, \citet{bellinger2024a} and \citet{henneco2024b} study the asteroseismic properties of more massive merger remnants that evolve through a long-lived BSG phase. Such stars have a convectively helium-burning core and a hydrogen-burning shell (\reffig{fig:kipp-mergers}), while single stars are almost entirely radiative (except for sub-surface convective layers). Because of these vastly different interior structures, they have very different gravity-mode period spacing patterns compared to single HG stars at the same position in the HR diagram, and the merged BSGs often also show deep dips. As in the lower-mass cases, the asymptotic period spacing pattern of merged BSGs is lower than in single stars. At the moment, however, it is not clear whether the predicted oscillation modes in BSGs are observable because the excitation mechanism(s) as well as the intrinsic amplitudes remain unclear. Observations of BSGs show low-frequency, stochastic variations and no regular modes have been identified so far \citep[see \eg][]{bowman2019a, bowman2024a, ma2024a}. Moreover, mode suppression by merger-produced magnetic fields has not been accounted for and could affect modes in the deeper g-mode cavities \citep{henneco2024b}.

In massive MS+MS star mergers, the merged stars can have unusual interior chemical gradients caused by a transient convective core and generally have a higher mean helium abundance in their envelopes. These structural differences, which are not present in single stars, also lead to characteristic oscillation properties in both gravity and pressure modes; however, it is much more challenging to observationally distinguish between merged stars and genuine single stars, and, \eg highly precise and accurate positions of stars in the HR diagram would be needed \citep{henneco2025b}.

\subsection{Further merger candidates}

The above sections cover several classes and groups of stars considered to be likely merger products. Here, we describe a few further merger candidates and discuss the cases of the progenitor of SN~1987A and $\eta$~Carinae in more detail. 

The RSG $\alpha$~Orionis, better known as Betelgeuse, has been suggested to be rapidly rotating and is known to be a runaway star with a prominent bow shock \citep{uitenbroek1998a, harper2017a, kervella2018a}. This led \citet{chatzopoulos2020a} to consider the possibility that Betelgeuse obtained its fast rotation and nitrogen-enriched surface in a merger. The interpretation of fast rotation from resolved velocity maps of Betelgeuse's surface \citep{kervella2018a} has since been questioned and could also be due to convective fluid motions \citep{ma2024b}. Moreover, fast rotation might not be a sign of stellar mergers. Further investigations into this interesting star will help to better understand its evolutionary history.

Other runaway stars such as HD~93521 \citep{gies2022a}, $\zeta$~Puppis \citep{vanbeveren2009a, pauldrach2012a}, and the BSG TYC\,3159-6-1 with a bipolar nebula \citep{gvaramadze2014a} have also been suggested to be the result of stellar mergers. Even one of the most massive runaways known today, the ${\approx}\,90\,\msun$ O-type star VFTS~016 in 30~Doradus \citep{evans2010a}, might be a merger product given its apparent young age and far distance from its likely birth cluster, R136 \citep{schneider2018a}. The apparent youth in comparison to the longer kinematic age of HD~93521 also points to a rejuvenated star, as in the case of VFTS~016. A merger origin could explain these age discrepancies. The formation scenario of these runaways would then be that of a dynamically ejected merged star or a binary system that later merged, possibly enabled by the ejection process. Alternatively, they could also be rejuvenated mass gainers of past binary mass transfer that were either ejected by cluster dynamics or by the SN explosion of their former donor stars. This might be the case for the fast-rotating runaways HD~93521 and $\zeta$~Puppis, while TYC\,3159-6-1 with its nebula and VFTS~016 might indeed be merged stars.

\subsubsection{SN~1987A and its blue supergiant progenitor Sanduleak -69$^\circ$~202}\label{sec:sn1987a}

SN~1987A was the first naked-eye SN since Kepler's SN in 1604, and puzzled astronomers for quite some time \citep[for reviews on SN~1987A, see \eg][]{podsiadlowski1992c, podsiadlowski2007b, podsiadlowski2017b}. The SN progenitor, Sanduleak -69$^\circ$~202, was a BSG rather than an RSG as expected for hydrogen-rich SNe at the time. It is surrounded by a relatively young (${\approx}20,000\,\mathrm{yr}$) bipolar triple-ring nebula (\reffig{fig:merger-debris}) that is enriched in CNO-processed material and helium. The slow expansion of material near the inner ring of ${\approx}\,10\,\kms$ suggests that this material is launched from an object of RSG dimensions, implying that Sanduleak -69$^\circ$~202 recently transitioned from a post-core-helium-burning RSG to a BSG. Moreover, the expanding SN material showed significant polarisation, hinting at an asymmetric explosion, and the SN light curve was distinct from other hydrogen-rich SNe.

Based on the apparently asymmetric explosion, \citet{chevalier1989a} were the first to postulate a binary origin for this SN. \citet{hillebrandt1989a} and \citet{podsiadlowski1990a} then suggested a CE merger model in which helium dredged up by the merger makes an RSG evolve into a BSG just before its SN. Direct simulations of the core merger inside a CE by \citet{ivanova2002a} showed efficient mixing of helium into the envelope, and the mass gain in a merger further helped the RSG to become a BSG \citep[\cf \refsec{sec:evolution-mergers}; see also][]{menon2017a, urushibata2018a}. In a series of papers, \citet{morris2006a, morris2007a, morris2009a} showed that one can also understand the triple-ring nebula as the ejecta of a CE merger, and the SN explosions of BSGs are also found to reproduce the light curve of SN~1987A \citep[\eg][]{dessart2010a, dessart2019a, utrobin2015a, menon2019a, tsuna2025a}.

\subsubsection{$\eta$ Carinae}\label{sec:eta-car}

$\eta$~Carinae and its Homunculus nebula are probably some of the most enigmatic stellar systems in the Milky Way (\reffig{fig:merger-debris}). Its Great Eruption in the 1840s made it the second brightest star in the night sky \citep{frew2004a} and likely created the Homunculus nebula. It is a highly eccentric, $e\,{\approx}\,0.9$, binary in a wide $5.54\,\mathrm{yr}$ orbit \citep{damineli2000a, grant2020a}, and many attempts have been made to make sense of all its multifaceted and detailed observations. The reader is referred to \citet{smith2018a} and \citet{hirai2021a} for an overview of these observational constraints.

A promising scenario to explain the $\eta$~Carinae system is that of a merger in a triple star system \citep[][]{smith2018a, hirai2021a}. Here, we follow the set of simulations carried out by \citet{hirai2021a} that demonstrate the feasibility of such a model. In this model, an initially stable, hierarchical triple system of three massive stars (\eg $40\text{--}60\,\msun$) is rendered unstable and becomes chaotic once binary mass transfer in the inner binary has expanded its orbit. The now chaotic triple encounters close and grazing encounters during which some mass is ejected in confined directions dictated by the encounter dynamics. These `sprays' could produce the observed outer ejecta of $\eta$~Carinae \citep{kiminki2016a}. After several close encounters and chaotic swaps of the stars, two of them get into contact, lose mass through the L$_2$ point and form a rather slow equatorial outflow with a velocity on the order of the orbital velocity, \ie $\mathcal{O}(100\,\kms)$. The outer tertiary star can plunge through this outflow upon each periastron passage, which may have created the precursor eruptions seen in 1838 and 1843 \citep{smith2011d}. When the two stars finally merge, a fraction of the orbital energy of $\mathcal{O}(10^{50}\,\mathrm{erg})$ thermalises by shocks and drives an eruption and optical transient that is reminiscent of LRNe. Such an explosive event could have created the ejecta seen moving with ${>}\,10,000\,\kms$ \citep{smith2018a}. Because of the previously lost mass in the orbital plane and the overall oblate shape of the system at this stage, much of the dynamic ejecta leaves through the polar region, creating a bipolar explosion. The central merged star, still containing excess energy and angular momentum from the coalescence, drives a strong and fast super-Eddington wind that sweeps up the dynamic ejecta into a thin, bipolar shell. This thin shell is the almost hollow Homunculus nebula that we observe today. Some of the ejecta may catch up with earlier ejectiles (\eg the sprays), possibly creating the observed X-ray emission \citep{seward2001a}. Such a model can naturally explain the observationally inferred ejecta mass and kinetic energy of the Homunculus nebula of ${>}\,10\,\msun$ and ${\approx}\,10^{50}\,\mathrm{erg}$ \citep{smith2006a}.

There remain open questions. For example, the nature of the 1890 eruption that probably formed the `Little Homunculus', a smaller bipolar nebula inside the Homunculus, is unclear \citep{ishibashi2003a}. Its mass ($0.1\,\msun$) and kinetic energy ($10^{46.9}\,\mathrm{erg}$) are orders of magnitude smaller than those of the Great Eruption, suggesting a different energy source and origin \citep{smith2005a}.

%
% Approximate merger methods
%
\section{APPROXIMATE\ MERGER\ METHODS}\label{sec:approximate-merger-methods}

Stellar mergers are a complex, inherently three-dimensional problem. Computing 3D merger models is computationally expensive, requiring approximate, computationally less expensive methods for certain applications. To this end, several methods have been developed, and we highlight a few here. It is possible to mimic the merger process by simply accreting the mass of one star onto the other on a faster-than-thermal timescale \citep[for example, this method or a variant thereof has been applied by][]{justham2014a, renzo2020a, rui2021a, schneider2024a}. Following Archimedes' principle, one can also construct the chemical structure of a merger remnant by entropy sorting \citep[\cf \refsec{sec:physics-of-mergers} and see, \eg][]{heller2025a, patton2025a}. A more sophisticated method is that of \citet{gaburov2008a}, which combines the idea of entropy sorting with results from 3D SPH head-on collision simulations to include entropy generation by shock heating. An earlier and different version of this approach is that of \citet{lombardi2002a}. There are further approximate methods in the literature, for example, those of \citet{podsiadlowski1992a}, \citet{ivanova2002c} and \citet{podsiadlowski2017a}, and the derivatives thereof of \citet{menon2017a} and \citet{urushibata2018a} to model the progenitor of SN~1987A. Neither of these methods can capture the full physics of stellar mergers, and the choice of the method depends on the application. We provide the Python package \texttt{PyStellarMerger} with implementations of the entropy-sorting technique and the \texttt{Make Me A Massive Star} (MMAMS) method of \citet{gaburov2008a} at \url{https://github.com/HITS-SET/PyStellarMerger}. For details of the implementations and benchmarks, see \citet{heller2025a}.

% Summary points
\begin{summary}[SUMMARY POINTS]
\begin{enumerate}
\item Stellar mergers are related to a wide range of astrophysical phenomena. These include blue straggler stars, long-lived blue supergiants, very massive stars and black holes, luminous red novae, bipolar nebulae, magnetic stars and highly-magnetic compact remnants, the large SN diversity, and many more.
\item There are multiple ways for stars to merge, \eg as a result of binary star evolution and by collisions in dense stellar systems. Stellar mergers occur frequently: ${>}\,15\text{--}40\%$ of binary stars may merge, while a similar fraction evolve through common-envelope phases. 
\item The merger process can be divided into four main phases: (i) contact/grazing encounters, (ii) coalescence, (iii) dynamical \& viscous, and (iv) thermal relaxation. During the dynamical/viscous relaxation, a central merger remnant coevolves with a rotationally-supported disc-like structure that likely regulates the final angular momentum budget of merged stars. The merger ejecta form bipolar nebulae.
\item Strong magnetic fields are produced in the merger process. Merged stars are thus likely responsible for some of the magnetic OBA stars and their possibly highly magnetic white-dwarf and neutron-star remnants, so-called polars and magnetars.
\item Initially, merger products rotate rapidly, but they can spin down efficiently and might then be observed as rather slowly spinning stars.
\item Merger products include rejuvenated main-sequence stars (blue stragglers and lurkers), long-lived blue supergiants, cool (super)giants with unusual core-to-envelope mass ratios, and their evolved descendants. Merger-produced blue supergiants are particularly distinct from single stars and, for example, give rise to a large variety of transients (\eg interacting and superluminous supernovae, and ultra-long gamma-ray bursts) and may collapse into very massive black holes that could even populate the pair-instability-supernova mass gap.
\item Transients connected to stellar mergers are luminous red novae from the merging process, and the merged stars contribute greatly to the large supernova diversity. The Great Eruption of $\eta$~Carinae and the luminous red nova V1309~Sco are examples of the former, while SN~1987A exemplifies the latter. 
\end{enumerate}
\end{summary}

% Future Issues
\begin{issues}[FUTURE ISSUES]
\begin{enumerate}
\item The initial conditions of binary mergers are crucial, yet they are not well understood. This applies to both the binary evolution leading up to contact (\eg binary mass transfer efficiency) as well as the physics of the preceding contact phases. 
\item The amount of mixing in mergers is still uncertain. In particular, this applies to mergers of post-MS and MS stars. For example, the mixing determines whether the merged star becomes a long-lived blue or red supergiant. No complete simulation of the merger of evolved stars has been accomplished yet. Resolving the dense and hot cores of such stars together with their low-density and rather cool envelopes is computationally prohibitive. Tiny timesteps are required for the core, while one has to evolve the model over a few comparably long, initial orbital periods.
\item While the mass loss of stellar collisions has been studied, systematic studies are lacking for binary mergers. For both stellar collisions and binary mergers, mass loss in the immediate merger aftermath (\eg via magnetically-driven outflows and disc winds) as well as during the thermal relaxation of the merged stars remains poorly understood.
\item The phase right after coalescence, when the central merger remnant accretes from a disc-like structure, is not well understood and warrants future investigations.
\item Magnetic fields are efficiently produced in stellar mergers, but it remains unclear for how long these fields can persist and what their surface configurations in fully relaxed merged stars look like. Because of this, it also remains uncertain how much these magnetic fields influence the evolution of merged stars, \eg via angular momentum transport and magnetic braking. 
\item While more and more theoretical and observational evidence points to merged stars being rather slow rotators, there remain significant uncertainties. Do all merged stars rotate rather slowly? Can some also rotate rapidly? What are the dominant physical processes that determine this? 
\item Observationally, it is often difficult to distinguish between merged stars and accretors of binary mass transfer. Investigating the binary nature of such stars would help greatly to establish the observed properties of merger products.
\end{enumerate}
\end{issues}

%Disclosure
\section*{DISCLOSURE STATEMENT}
The authors are not aware of any affiliations, memberships, funding, or financial holdings that
might be perceived as affecting the objectivity of this review.

% Acknowledgements
\section*{ACKNOWLEDGMENTS}
I am extremely grateful to Jan Henneco and Pablo Marchant for their helpful feedback on this manuscript, and to Sebastian Ohlmann and Max Heller for their help in preparing figures for this review and the approximate merger methods. Many thanks also to Philipp Podsiadlowski, R{\"u}diger Pakmor, and the astrophysics groups at HITS for the many fruitful discussions on stellar mergers and common envelope evolution. 
I gratefully acknowledge support from the Klaus Tschira Foundation. This work has received funding from the European Research Council (ERC) under the European Union's Horizon 2020 research and innovation programme (Grant agreement No.\ 945806) and is supported by the Deutsche Forschungsgemeinschaft (DFG, German Research Foundation) under Germany's Excellence Strategy EXC 2181/1-390900948 (the Heidelberg STRUCTURES Excellence Cluster).

% References
\bibliographystyle{ar-style2.bst}

\end{document}